\documentclass[12pt]{elsart}
\usepackage{graphicx}
\usepackage{amsmath}
\def\om{\omega}
\def\rmF{{\rm F}}
\def\rmd{{\rm d}}
\newcommand{\lsim}{\stackrel{\scriptstyle <}{\phantom{}_{\sim}}}
\newcommand{\gsim}{\stackrel{\scriptstyle >}{\phantom{}_{\sim}}}

\begin{document}
	
\begin{frontmatter}
\title{Delta isobars in relativistic mean-field  models with $\sigma$-scaled hadron masses and couplings}
\author[UMB]{E.E.~Kolomeitsev},
\author[MEPHI]{K.A.~Maslov} \and
\author[MEPHI]{D.N.~Voskresensky}
\address[UMB]{Matej Bel University, SK-97401 Banska Bystrica, Slovakia}
\address[MEPHI]{National Research Nuclear University ``MEPhI'', RU-115409 Moscow, Russia}
\begin{abstract}
We extend the relativistic mean-field models with hadron masses and meson-baryon coupling constants dependent on the scalar $\sigma$ field, studied previously to incorporate $\Delta(1232)$ baryons.
Available empirical information  is analyzed to put constraints on  the  couplings of $\Delta$s with meson fields.  Conditions for the appearance of $\Delta$s are studied.
We demonstrate that with  inclusion of the $\Delta$s our equations of state continue to fulfill majority of known empirical constraints including the pressure-density constraint from heavy-ion collisions,  the constraint on the maximum mass of the neutron stars, the direct Urca and the gravitational-baryon mass ratio constraints.
\end{abstract}
\end{frontmatter}

\tableofcontents

\section{Introduction}

A nuclear equation of state (EoS) is the key ingredient in the description of neutron stars (NSs)~\cite{Lattimer:2012nd}, supernova explosions~\cite{Woosley} and heavy-ion
collisions~\cite{Danielewicz:2002pu,Fuchs}. Relativistic mean-field (RMF) models
are widely used for construction of a hadronic EoS. The original model~\cite{Durr,Walecka1974} included interaction of nucleons with scalar ($\sigma$) and vector ($\omega$) meson mean fields. Next, for the better description of the symmetry energy the isovector ($\rho$) meson field was incorporated and the work~\cite{Boguta77} included a $\sigma$-field self-interaction in the form of the potential $U(\sigma)=b\sigma^3/3+c\sigma^4/4$. The coupling constants $b$ and $c$ were adjusted to describe the saturation properties of the isospin symmetrical nuclear matter: the saturation  density $n_0$, the binding and symmetry energies, and the incompressibility coefficient at the nuclear saturation.

At present, there exists a vast number of modifications of the RMF models. They differ by extra terms in the effective Lagrangian related to new fields and their interactions, cf.~\cite{SerotWalecka,Glendenning,Weber,Savushkin2015} and references therein. Various experiments indicate  modifications of hadronic properties in hadronic matter~\cite{Metag}. To improve agreement between theoretical descriptions and experimental data RMF models with density-dependent meson-nucleon coupling constants were developed, cf.~\cite{Fuchs,Typel,Hofmann,Niksirc,Gaitanos,Long,Lalazisis,Typel2005,Voskresenskaya,RocaMaza:2011qe,
Dutra:2014qga,Dutra:2015hxa}. On the other hand, due to a partial restoration of the chiral symmetry in dense and/or hot matter masses of all hadrons except Goldstone bosons, like pions and kaons, are expected to decrease with increasing density and/or temperature, cf.~\cite{Rapp,Koch}. According to the conjecture of Brown and Rho~\cite{BrownRho} the nucleon mass and the masses of vector $\omega$, $\rho$ and scalar $\sigma$ mesons should obey an approximately universal scaling law.  Motivated by these ideas  two of us demonstrated in~\cite{Kolomeitsev:2004ff} how one can construct RMF models incorporating simultaneously in-medium modifications  of the baryon and meson masses and coupling constants. In~\cite{Kolomeitsev:2004ff} the effective hadron masses are assumed to be $\sigma$-field dependent. The density dependence of the $\sigma$ field can be related to a  modification of the chiral condensate in the medium. Also, in the lattice QCD in the strong coupling limit~\cite{Ohnishi:2008yk}  meson masses are approximately proportional to the equilibrium value of the chiral condensate, and the latter value  decreases with an increase of the baryon density. Remarkably, in the case of infinite matter the effective hadron masses ($m^{*}_h$) and the coupling constants ($g^{*}_h$) enter all relations only in  combinations $m^{*\,2}_h/g_h^{*\,2}$ that leads to equivalence between  different  RMF schemes~\cite{Kolomeitsev:2004ff}. Allowing for differences in scaling functions for hadron masses and coupling constants one can better fulfill various experimental constraints on the EoS.

A comparison between different nucleon EoSs in how well they satisfy various empirical constraints was performed in~\cite{Klahn:2006ir} for the EoSs obtained in the RMF models, in more microscopic approaches~\cite{APR,Gandolfi:2009nq}, and in Skyrme models~\cite{Dutra12}.
Some of the previously used constraints~\cite{Klahn:2006ir} were recently tightened and
new constraints were formulated. At present there exists an agreement that the EoS of the cold hadronic matter should:
($i$)
satisfy experimental information on properties of dilute nuclear matter
and not contradict results of microscopically based approaches;
($ii$)
fulfill empirical constraints extracted from the description of global characteristics of atomic nuclei, for the baryon density $n$ near the saturation nuclear matter density $n_0\simeq 0.16\,$fm$^{-3}$;
($iii$) not contradict constraints on the pressure of the nuclear mater at densities above $n_0$ extracted from the description of particle transverse and elliptic flows~\cite{Danielewicz:2002pu} and the $K^+$ production~\cite{Lynch} in heavy-ion collisions;
($iv$) allow for the heaviest known compact stars PSR~J1614-2230 with the mass $1.97\pm 0.04\,M_{\odot}$ and PSR~J0348+0432 with the mass $2.01\pm 0.04 \,M_{\odot}$~\cite{Demorest:2010bx,Antoniadis:2013pzd} ($M_\odot$ is the solar mass);
($v$) allow for an adequate description of the compact star cooling, which is possible, if the most efficient direct Urca (DU) neutrino processes $n\to p+e+\bar{\nu}_e$, $p+e\to n+\nu_e$ do not occur in the majority of the known pulsars~\cite{Kolomeitsev:2004ff,Blaschke:2004vq,Grigorian:2016leu}\footnote{The problem with the large contribution from nucleon DU reaction to NS cooling can be avoided, if one uses very large neutron or proton pairing gaps~\cite{Taranto-DU}. };
($vi$) explain the gravitational mass and total baryon number of pulsar PSR J0737-3039(B) with at most 1\% deviation from the baryon number predicted for this object~\cite{Podsiadlowski,Kitaura:2005bt};
($vii$) yield a mass-radius relation comparable with the empirical constraints~\cite{Bogdanov:2012md,Hambaryan2014,Heinke:2014xaa};
($viii$) being extended to non-zero temperatures,  appropriately describe heavy-ion collision data.

Analysis performed in many papers demonstrated that it is most difficult to reconcile the constraint on the maximum NS mass, $1.97\,M_\odot$, cf.~\cite{Demorest:2010bx,Antoniadis:2013pzd}, and the constraints on the stiffness of the EoS extracted from the analyses of the flow in heavy-ion collisions~\cite{Danielewicz:2002pu,Fuchs}.

In~\cite{Kolomeitsev:2004ff} the model MW(n.u., $z=0.65$), labeled in~\cite{Klahn:2006ir}
as the KVOR model, was constructed. As shown in~\cite{Klahn:2006ir} the KVOR model allowed to satisfy appropriately the majority of experimental constraints known to that time including the flow constraint. In~\cite{Khvorostukhin:2006ih,Khvorostukhin:2008xn} the model was extended to finite temperatures and successfully applied to the description of heavy-ion collisions. However, the KVOR EoS  supplemented by  the
Baym-Pethick-Sutherland EoS for the NS crust~\cite{Baym:1971pw} yields $M_{\rm max}[{\rm KVOR}]=2.01\,M_{\odot}$ that fits the constraint~\cite{Demorest:2010bx,Antoniadis:2013pzd} only marginally. A possibility of the population of hyperon Fermi seas in dense beta-equilibrium matter (BEM) was not incorporated. The problems with the EoS worsen, however, when strangeness is included, because the appearance of hyperons leads to a softening of the EoS and to the reduction of the maximum NS mass. It  is possible to explain observed  massive NSs only if one artificially forbids the appearance of hyperons that cannot be reconciled with the known information on  binding energies of hyperons in nuclear matter extracted from hypernuclei, see~\cite{Djapo:2008au,Glendenning} and references therein. This is called the ``hyperon puzzle''. The difference between NS masses with and without hyperons proves to be so large for the reasonable choices of hyperon coupling constants in the standard RMF approach, that in order to solve the puzzle one has to start with a very stiff nucleon EoS that hardly agrees with the results of microscopically-based calculations using the variational~\cite{APR} and auxiliary-field diffusion Monte Carlo~\cite{Gandolfi:2009nq} methods. Such an EoS would also be incompatible with the restrictions on the EoS stiffness extracted from the analyses of the particle flows in heavy-ion
collisions~\cite{Danielewicz:2002pu,Fuchs}. All suggested explanations require additional assumptions, see discussion in~\cite{Fortin:2014mya}.

In recent papers~\cite{Maslov:2015msa,Maslov:2015wba} we proposed two modifications of the KVOR model~\cite{Kolomeitsev:2004ff}. One extension of the model (KVORcut) demonstrates that the EoS stiffens, if a growth of the scalar-field magnitude with an increase of the density is bounded from above at some value for baryon densities exceeding a certain value above $n_0$. This can be realized,  if the nucleon -- vector-meson coupling constant changes rapidly as a function of the scalar field slightly above the desired value. The other version of the model (MKVOR) assumes a smaller value of the nucleon effective mass at the nuclear saturation density and uses a saturation of the scalar field in the isospin asymmetric matter induced by a strong variation of the nucleon --  isovector-meson coupling constant as a function of the scalar field. A possibility of hyperonization of the matter in NS interiors was taken into account. The resulting EoSs fulfill a majority of known empirical constraints including the pressure-density constraint from heavy-ion collisions, direct Urca constraint, gravitational-baryon mass constraint for the pulsar J0737-3039B, and the constraint on the maximum mass of the NSs.

Similar problem may arise if  new baryon species are incorporated in RMF models. The next in the mass order are $\Delta(1232)$ isobars. Their appearance in NS interiors may lead to similar effects as for hyperons. In~\cite{Maslov:2015msa,Maslov:2015wba} the $\Delta$ isobars were not included.

The $\Delta$ baryons play the very important role in nuclear physics~\cite{Cattapan02}.
They contribute essentially to the pion polarization operator in the nuclear medium leading to an enhancement of the pion softening with an increase of the baryon density and promoting thereby a pion condensation at nucleon densities above a critical density, $n>n_c^{\pi}>n_0$, cf.~\cite{Migdal78,EricsonWeise,Migdal:1990vm}. With some assumptions about $\pi N\Delta$ and/or $\Delta\Delta \sigma$ interactions in dense nucleon matter one speculated in~\cite{Migdal78,Boguta1982} about a possibility of density isomer states.
Also, $\Delta$s are produced copiously in energetic heavy-ion collisions~\cite{Metag} and their in-medium modifications may lead to important observable consequences~\cite{Cubero:1987pr,Voskresensky:1993ud,Khvorostukhin:2006ih,Khvorostukhin:2008xn}.

 During a long time the presence of $\Delta$ baryons in NSs was regarded as an important but unresolved issue~\cite{ShapiroTeukolsky,Sawyer72}. In the RMF model, in which $\Delta$s couple to meson fields with the same strength as nucleons~\cite{Glendenning} the critical density for the appearance of $\Delta$ isobars was estimated as $\sim 10 n_0$. Therefore, implying that in BEM the critical density for the appearance of $\Delta$s should be also high, one devoted much less effort to the study of $\Delta$ baryons in NSs compared to the investigation of the
possible appearance of hyperons.
%
%
The issue was reconsidered in~\cite{Xiang,Chen,Schurhoff,Lavagno} and more recently in~\cite{Drago2014,Cai:2015hya,Drago:2015cea}. Using different density dependencies of the nuclear symmetry energy and assumptions about the baryon-meson coupling constants the authors came to conclusion about feasible effects of $\Delta$ on both the composition and structure of NSs. References~\cite{Drago2014,Drago:2015cea} formulated the problem as the ``$\Delta$ puzzle", which could exist on equal footing with the hyperon puzzle.


In this work we include $\Delta$ resonances in the RMF models with scaled hadron masses and couplings --- KVORcut03 and MKVOR --- suggested recently in~\cite{Maslov:2015msa,Maslov:2015wba}. In the absence of $\Delta$s these models have appropriately passed  mentioned above constraints. We analyze, if within these models one is able  to construct the appropriate EoS with hyperons and $\Delta$ baryons, satisfying presently known experimental constraints.

Our work is organized as follows. In Section~\ref{sec:model} we formulate our generalized RMF model with $\sigma$-field scaled hadron masses and couplings with inclusion of  $\Delta$ isobars. In Section~\ref{sec:eos} we first investigate KVORcut03 and MKVOR models with $\Delta$ baryons (i.e., KVORcut03$\Delta$ and MKVOR$\Delta$ models).
We show that in the MKVOR$\Delta$ model the effective nucleon mass in isospin-symmetric matter (ISM) drops to zero at $n\sim (4-6)n_0$, if one exploits a relevant value for the $\Delta$ potential $U_\Delta \sim -(50-100)$~MeV. Then within this model for a higher density the hadronic EoS cannot be used and should be replaced to the quark one. In order continue to deal with the hadron description we slightly modify the MKVOR model and label it as MKVOR*. Results of numerical calculations are presented in Section \ref{Numerical}. We demonstrate that within the KVORcut03-based and MKVOR*-based models one is still able  to construct the appropriate EoS with inclusion of hyperons and $\Delta$s, satisfying presently known experimental constraints. Our final results are summarized in the Conclusion.

\section{Lagrangian and energy-density}\label{sec:model}

\subsection{RMF model with scaled hadron masses and couplings. General formalism}

We will closely follow the approach described in~\cite{Kolomeitsev:2004ff,Maslov:2015msa,Maslov:2015wba} and include now besides the full SU(3) ground-state baryon octet also the isospin quadruplet of $\Delta$ baryons $\Delta = (\Delta^-, \Delta^0, \Delta^+, \Delta^{++})$.
In the mean-field approximation we can disregard all complications related to the structure of the wave function of the spin-$3/2$ baryons and treat $\Delta$ as  spin-$1/2$ fermions with the bare mass $m_\Delta=1232$~MeV and the spin degeneracy factor 4.
Baryons  $b = (n,p,$ $\Lambda,$ $\Sigma^{\pm,0},$ $\Xi^{-,0}; \Delta^{\pm,0,++})$ interact with meson mean fields, $m=(\sigma, \omega, \rho,\phi)$,  $\sigma$ is  the scalar meson  and $\omega, \rho,\phi$ are vector mesons. The baryon contribution to the Lagrangian density is
\begin{align}
\mathcal{L}_{\rm bar}&=
\sum_{b} \bar{\Psi}_{b}\big[i \FMslash{D} -m_b\Phi_b\big] \Psi_b, \quad \FMslash{D}=\gamma^{\mu}D_\mu\,, \\ D_\mu &= \partial_\mu + i g_{\om b} \chi_{\om b} \omega_\mu +  i g_{\rho b} \chi_{\rho b}\vec{t}_b \vec{\rho}_\mu + i g_{\phi b} \chi_{\phi b} \phi_\mu\,.\nonumber
\label{Lag-bar}
\end{align}
Here $\Psi_{b}$ stands for the bispinor of the spin-$1/2$ baryon (and symbolically for
the Rarita-Schwinger spinor with contracted indices for spin-$3/2$ particle). The summation runs over all twelve baryonic states $b$; $\gamma^\mu$ are Dirac $\gamma$-matrices, $\vec{t}_b$ is the baryon isospin operator, which projection is expressed through the baryon electric charge $Q_b$ and strangeness $S_b$ as
$t_{3b}=-\frac12+Q_b-\frac12 S_b$ (recall $S_{N,\Delta}=0$, $S_{\Lambda,\Sigma}=-1$ and $S_{\Xi}=-2$).
The meson field contribution  to the Lagrangian density is
\begin{eqnarray}
\mathcal{L}_{\rm mes}  &=& \half \partial_\mu \sigma \partial^\mu \sigma
- \half m_\sigma^{2}\Phi_\sigma^2 \sigma^2  - {U}
-\quart\om_{\mu \nu} \om^{\mu \nu} + \half m_\omega^{2}\Phi_\omega^2\, \om_\mu \om^\mu
\nonumber \\
& -& \quart\vec{\rho}_{\mu \nu} \vec{\rho}\,^{\mu \nu}
+ \half m_\rho^{2}\Phi_\rho^2 \vec{\rho}_\mu \vec{\rho}^{\,\mu}
- \quart\phi_{\mu \nu} \phi^{\mu \nu}+ \half m_\phi^{2}\Phi_\phi^2\, \phi_\mu \phi^\mu
,
\label{Lag-mes}
\end{eqnarray}
where $\om_{\mu \nu} = \partial_\mu \om_\nu - \partial_\nu \om_\mu$,
$\phi_{\mu \nu} = \partial_\mu \phi_\nu - \partial_\nu \phi_\mu$
and for the $\rho$ meson we  take into account self-interaction via the non-Abelian long derivative $\vec{\rho}_{\mu	\nu} = \partial_\mu \vec{\rho}_\nu -  \partial_\nu \vec{\rho}_\mu + g_\rho' \chi_\rho' [\vec \rho_\mu \times \vec \rho_\nu]$. The latter term proves to be important in the discussion of a charged $\rho$ condensation proposed in~\cite{v97,Kolomeitsev:2004ff}. In the given work we suppress this possibility.

Within the approach of Ref.~\cite{Kolomeitsev:2004ff} the
effective coupling constants in matter  depend on the $\sigma$ field  via the scaling functions as $g_{\sigma b}^*=g_{\sigma b} \chi_{\sigma b}(\sigma)$, $g_{\omega b}^* =g_{\omega b} \chi_{\omega b}(\sigma)$, $g_{\rho b}^* =g_{\rho b} \chi_{\rho b}(\sigma)$, $ g_{\phi b}^* =g_{\phi b}
\chi_{\phi b}(\sigma)$, $g_\rho^{'*} =g_\rho'\chi_\rho'(\sigma)$. The potential
$U(\sigma)$ allows for a self-interaction of the $\sigma$ field.
In matter the bare masses of baryons, $m_b$, and mesons, $m_{m}$, are replaced by the effective masses $m_b^*=m_b\Phi_b(\sigma)$, $m_{m}^*=m_m\Phi_m(\sigma)$.

The full Lagrangian density of the model is  given by the sum $\mathcal{L} = \mathcal{L}_{\rm bar}
+ \mathcal{L}_{\rm mes} + \mathcal{L}_{\rm lept}$, where to describe the BEM  we also include the Lagrangian density of light leptons: electrons and muons, $\mathcal{L}_{\rm lept} = \sum_l \bar{\psi}_l (i \partial_\mu \gamma^\mu - m_l) \psi_l$, $l=e,\mu$;  $\psi_l$ stands  for the lepton bispinor and $m_l$ is the bare lepton mass.
Masses of all particles are taken the same as in~\cite{Maslov:2015wba,Kolomeitsev:2004ff}.

The $\sigma$ field dependence enters the scaling function $\chi_{mb}$ and $\Phi_{b(m)}$ through an auxiliary variable
\begin{align}
f = g_{\sigma N} \chi_{\sigma N} (\sigma) \frac{\sigma}{m_N}\,.
\end{align}
As in \cite{Kolomeitsev:2004ff,Khvorostukhin:2006ih,Khvorostukhin:2008xn,Maslov:2015msa,Maslov:2015wba} we  exploit the universal scaling functions for the nucleon and meson masses:
\begin{align}
\Phi_N (f)=\Phi_m (f)=1-f,
\label{PhiN}
\end{align}
but allow for a variation of the scaling functions of coupling constants $\chi_{m b}$.
We suppose that $\chi_{\omega b}(f) = \chi_{\omega N}(f)$\,, $\chi_{\rho b}(f)
= \chi_{\rho N}(f)$. Then the scaling function $\Phi_b$ for all baryons including hyperons and $\Delta$ isobars
can be written as
\begin{align}
\Phi_b (f)=\Phi_N\big(g_{\sigma b}\chi_{\sigma b}\frac{\sigma}{m_b}\big)\equiv
\Phi_N\big(x_{\sigma b}\xi_{\sigma b}\, \frac{m_N}{m_b}\,f\big)\,,
\quad \xi_{\sigma b}=\frac{\chi_{\sigma b}}{\chi_{\sigma N}}\,,
\end{align}
where $\xi_{\sigma b}$ is a function of $f$.

 With the help of equations of motion for vector fields, in the standard way we recover the energy-density functional for the cold infinite baryonic matter of an arbitrary isospin composition~\cite{Glendenning,Weber,Kolomeitsev:2004ff,Maslov:2015wba}:
\begin{eqnarray}
E[\{n_b\};f] &=& \sum_{b } (2 s_b+1) E_{\rm kin}(p_{\rmF,b},m_b^*(f)) + \sum_{l=e,\mu} 2E_{\rm kin}(p_{\rmF l}, m_l)\nonumber\\
& +& \frac{m_N^4 f^2}{2 C_\sigma^2}  \eta_\sigma(f)
+ \frac{1}{2 m_N^2} \Big[\frac{C_\om^2 n_B^2}{ \eta_\om(f)}
+\frac{C_\rho^2 n_I^2}{\eta_\rho(f)}
+ \frac{C_\phi^2 n_S^2}{\eta_\phi(f)}
\Big]\,,
\label{En}\\
C_M &=& g_{MN} \frac{m_N}{m_M}\,,\,\,M=(\sigma,\om,\rho)\,,\quad C_\phi=C_\om \frac{m_\om}{m_\phi}\,,
\nonumber \end{eqnarray}
where $s_b$ stands for the fermion spin. The fermion energy is given by
\begin{eqnarray}
E_{\rm kin}(m,p_\rmF) &=&\frac{1}{16 \pi ^2}\left(p_\rmF \sqrt{m^2 + p_\rmF^2} (m^2+2 p_\rmF^2)-m^4 {\rm arcsinh}(p_\rmF/m)  \right)\,,
\nonumber
\end{eqnarray}
with the Fermi momentum of  species $b$ related to the number density as  $p_{\rmF,b}=(6\pi^2\,n_b/(2s_b+1))^{1/3}$.
In Eq.~(\ref{En}) we introduced effective densities of baryon number, isospin and strangeness,
\begin{eqnarray}
&& n_B = \sum_b x_{\om b} n_b, \quad n_I = \sum_b x_{\rho b} t_{3b} n_b, \quad
n_S = \sum_b x_{\phi b} n_b, \,
\nonumber\\
&& \mbox{with}\quad x_{\om b} = \frac{g_{\om b}}{g_{\om N}}\,,\quad
x_{\rho b} = \frac{g_{\rho b}}{g_{\rho N}}\,,\quad
x_{\phi b} = \frac{g_{\phi b}}{g_{\om N}},
\label{nBIS}
\end{eqnarray}
which determine the contributions from mean fields of the vector mesons to the total energy density.

The key difference of our approach from the standard non-linear Walecka-like RMF models is the presence of the scaling functions for the vector meson fields $\eta_{\om,\rho,\phi}$, which stand for the ratios of the scaling functions for the hadron mass and the coupling constant
\begin{eqnarray}
\eta_\om (f) = \frac{\Phi_\om^2 (f)}{\chi_{\om N}^2 (f)}\,, \quad \eta_\rho (f) = \frac{\Phi_\rho^2 (f)}{\chi_{\rho N}^2 (f)}\,, \quad  \eta_\phi (f) = \frac{\Phi_\phi^2 (f)}{\chi_{\phi N}^2 (f)}\,.
\label{eta-def}
\end{eqnarray}
We stress that, as long as we consider an infinite system, there is actually no need to specify the scaling functions $\Phi_\om$, $\chi_\om$, $\Phi_\rho$, $\chi_\rho$, $\Phi_\phi$, and $\chi_\phi$ separately, but only their combinations \cite{Kolomeitsev:2004ff}.

 The scalar-field self-interaction potential ${U}(\sigma)$ can be   hidden in the scaling function $\eta_{\sigma}(f)$ that we further assume:
\begin{eqnarray}
\eta_{\sigma}(f)=\frac{\Phi_{\sigma}^2[\sigma(f)]}{\chi_{\sigma N}^2[\sigma(f)]} + \frac{ 2 \, C_{\sigma}^2}{m_N^4 f^2}  {U}[\sigma(f)]\,.
\label{Uinetaf}
\end{eqnarray}

The equation of motion for the remaining field variable $f$ follows from the minimization of the energy density (\ref{En}), 
\begin{eqnarray}
\frac{m_N^3\,f}{C_\sigma^2 } \eta_\sigma(f)&=&  n_{B,\rm sc}(f,\{n_b\}) + n_{\rm MF}(f,\{n_b\})\,,
\label{eq_fn}
\end{eqnarray}
where the source of the scalar field is now not only the baryon scalar density
\begin{eqnarray}
n_{B,\rm sc}(f,\{n_b\}) =  -\sum_{b} \frac{m_b}{m_N} \Phi'_b(f)(2s_b+1) \rho_{\rm sc} (m_b \Phi_b(f), p_{\rmF,b}),
\nonumber\\
\rho_{\rm sc}(m,p_\rmF) =
\frac{1}{4 \pi ^2} \big(m p_\rmF \sqrt{m^2 + p_\rmF^2}- m^3 {\rm arcsinh}(p_\rmF/m)\big)   \,,
\label{rhoS}
\end{eqnarray}
but also meson contributions due to the mean-field scaling functions
\begin{align}
n_{\rm MF}(f,\{ n_b \})
=  \frac{C_\om^2 \eta'_\om (f)n_B^2}{2m_N^3\eta^2_\om(f)}
+ \frac{C_\rho^2 \eta'_\rho(f)n_I^2}{2m_N^3\eta^2_\rho(f)}
+ \frac{C_\phi^2 \eta'_\phi(f)n_S^2}{2m_N^3\eta^2_\phi(f)}
-\frac{m_N^3 f^2}{2 C_\sigma^2 } \eta_\sigma'(f).
\label{rho-eta}
\end{align}

The chemical potential for the baryon species $b$ can be calculated as $\mu_b=
\frac{\partial}{\partial n_b} E[\bar{f},\{n_b\}]$, where $\bar{f}$ is the solution of Eq.~(\ref{eq_fn}) for given partial densities of baryons $\{n_b\}$, or explicitly
\begin{eqnarray}
\mu_b=\sqrt{p_{\rmF,b}^2+m_b^2\Phi_b^2(\bar{f})}
+\frac{1}{m_N^2} \Big[x_{\om b}\frac{C_\om^2 n_B}{ \eta_\om(\bar{f})}
+x_{\rho b}t_{3b}\,\frac{C_\rho^2 n_I}{\eta_\rho(\bar{f})}
+ x_{\phi b}\frac{C_\phi^2 n_S}{\eta_\phi(\bar{f})}
\Big].
\label{mub}
\end{eqnarray}

The composition is determined by the conditions of chemical equilibrium with respect to the processes which can occur in the medium. If we consider the nuclear matter on a short time-scale, so that weak processes have no time to occur, hyperons cannot appear but the $\Delta$ admixture can be created and balanced by fast strong processes $N N \leftrightarrow \Delta N$ and $NN\leftrightarrow \Delta\Delta$. The latter ones impose the relations among chemical potentials
\begin{eqnarray}
\label{eq4muD}
\mu_{\Delta^-}=2\mu_n-\mu_p\,,\quad \mu_{\Delta^0}=\mu_n\,,\quad \mu_{\Delta^+}=\mu_p\,,
\quad \mu_{\Delta^{++}}=2\mu_p -\mu_n\,,
\end{eqnarray}
where the nucleon chemical potentials, $\mu_n$ and $\mu_p$, are fixed by the total baryon, $n_B$, and isospin, $n_I$, densities. These conditions will be used to determine the $\Delta$ amount in the ISM, which is defined by the condition $n_I=0$, and therefore $\mu_n=\mu_p$.
In a long-living system like a NS the weak processes have enough time to occur and we deal with  the BEM. Thus, the composition of the NS core is determined by conditions of the $\beta$-equilibrium, which impose the relations among the particle chemical potentials
\begin{align}
\mu_b=\mu_n - Q_b\,\mu_e \,,\quad \mu_e =\mu_\mu,
\label{chempot-i}
\end{align}
and by the electro-neutrality condition
\begin{align}
\sum_{b} Q_b\,n_b-n_e-n_\mu=0\,,
\label{electroneut}
\end{align}
where lepton densities $n_l$ are given by $n_l=(\mu_l^2-m_l^2)^{3/2}/(3\pi^2)$\,, $l=e,\mu$.
Solving Eqs.~(\ref{chempot-i}) and (\ref{electroneut}), one can obtain the particle densities $n_i$, $i=(b,l)$, as functions of the total baryon density $n_B \equiv n= \sum_{b} n_b$. Finally, pressure of the matter in the $\beta$-equilibrium can be calculated as
\begin{align}
P[n, n_i]=\sum_i\mu_i\, n_i -E[\bar{f}(n),\{n_i\}]\,.
\label{press}
\end{align}
The sum runs here over all baryons and leptons. For densities $n< 0.7 n_0$ we match the RMF EoS with the BPS crust EoS, see Appendix A in~\cite{Maslov:2015wba} for details. The final NS configuration follows from the solution of the Tolman--Oppenheimer--Volkoff equation.

Now, it remains to specify the ratios of the coupling constants in~(\ref{nBIS}).

\subsection{Couplings for hyperons}

The coupling constants of hyperons to vector mesons can be related to those of nucleons with the help of the SU(6) symmetry relations:
\begin{align}
&x_{\om \Lambda} = x_{\om \Sigma}= 2 x_{\om \Xi}=\frac{2}{3} \,,
\quad
x_{\rho\Lambda}=0\,, \quad  x_{\rho \Sigma} = 2x_{\rho \Xi} = 2\,, \nonumber\\
&x_{\phi \Lambda} =  x_{\phi \Sigma} = \frac12 x_{\phi \Xi} = -\frac{\sqrt{2}}{3} \,,
\quad x_{\phi N} = 0.
\label{gHm}
\end{align}
The scalar meson coupling constants are constrained by hyperon potentials, $U_H(n_0)$, or, equivalently, by the hyperon binding energies in the nucleon ISM at saturation, which are deduced from extrapolations of hyper-nucleus data,
\begin{eqnarray}
x_{\sigma H}=\frac{x_{\omega H} n_0 C_{\omega}^2\eta_\om(\bar{f}_0)/{m_N^2}
-U_{H}(n_0)}{m_N-m_N^{*} (n_0)}\,,
\label{EHbind}
\end{eqnarray}
where we put $\xi_{\sigma H}(\bar{f}_0)=1$, and $\bar{f}_0$ is the solution of equation of motion in the ISM at saturation, $n_p=n_n=n_0/2$. Note that the $\eta_\om$ scaling will be chosen later so that $\eta_\om(\bar{f}_0)\approx 1$. As in~\cite{Maslov:2015msa,Maslov:2015wba} we will use the values
$$U_{\Lambda }(n_0) = -28 \,{\rm MeV},\quad U_{\Sigma }(n_0) = 30 \,{\rm MeV},\quad U_{\Xi }(n_0) = -15 \,{\rm MeV}\,.$$

The described scheme  leaves us a freedom for choosing the scaling functions $\eta_\phi(f)$ and $\xi_{\sigma H}(f)$. Following \cite{Maslov:2015msa,Maslov:2015wba}, we consider two choices. The first choice (which we  label by H$\phi$ suffix) is when we incorporate the $\phi$-meson mean field with the very same scaling of the $\phi$ mass as for all other hadrons, $\Phi_\phi=1-f$, but use unscaled coupling constants $\chi_{\phi b}=1$,
\begin{eqnarray}
\eta_\phi=(1-f)^2\,,\quad\mbox{and}\quad \xi_{\sigma H}=1\,.
\label{hyp-Hphi}
\end{eqnarray}
In the second  choice (labeled by H$\phi\sigma$ suffix) we use
\begin{eqnarray}
\eta_\phi=(1-f)^2\,,\quad\mbox{and}\quad \xi_{\sigma H}=
\left\{\begin{array}{cc}
1\,,  \quad\mbox{for} \quad  n = n_0\\
0\,,  \quad\mbox{for} \quad  n \geq n_{cH}\end{array}\right.
\,,
\label{hyp-Hphisig}
\end{eqnarray}
where $n_{cH}$ is the critical density for hyperonization. With this  assumption $\xi_{\sigma H}$  decreases reaching zero for the baryon density  $n = n_{cH}$ and for $n \geq n_{cH}$ we  exploit vacuum masses for the hyperons. Note that the KVOR model extended to the high temperature regime in Ref. \cite{Khvorostukhin:2008xn} (the SHMC model) matches well the lattice data up to temperature 250 MeV provided all the baryon-$\sigma$ coupling constants except the nucleon ones are artificially suppressed, that partially motivates our second choice of $\xi_{\sigma \rm H}=0$ for densities at which the hyperons are produced.
Introducing the scalings (\ref{hyp-Hphi}), (\ref{hyp-Hphisig}) allowed us to resolve the hyperon puzzle within our models \cite{Maslov:2015msa,Maslov:2015wba}.


\subsection{Couplings for $\Delta$ baryons}

The coupling constants of the $\Delta$ resonances are poorly constrained empirically, due to  unstable nature of the $\Delta$ particles and  the complicated pion-nucleon  dynamics in-medium. Simplest is the universal choice of the couplings of the $\Delta$ with $\sigma$, $\om$, $\rho$ fields, which is usually argued by a naive quark counting~\cite{Glendenning}:
\begin{eqnarray}
x_{\om\Delta}=x_{\rho\Delta}=x_{\sigma\Delta}=1\,, \quad x_{\phi\Delta}=0.
\label{x-QC}
\end{eqnarray}

The range of possible deviations from the universal law  was investigated in \cite{SerotWalecka,Wehrberger1,Kosov,Oliveira,Zschiesche}, see also \cite{Glendenning}.
The choice of coupling parameters (\ref{x-QC}) in $\sigma$ and $\om$ sectors  assumes that potentials acting on $\Delta$ and nucleons are the same. There are, however, experimental evidences that these potentials can be essentially different already in the ISM. To allow for a deviation from the universal scaling we, similarly to the hyperon case, cf. Eq.~(\ref{EHbind}),  include an
additional constraint on the $x_{\sigma\Delta}$ from the potential of the $\Delta$ baryon, $U_\Delta(n_0)$, in the ISM at saturation density $n_0$:
\begin{eqnarray}
x_{\sigma \Delta}=\frac{x_{\omega \Delta}C_{\omega}^2 n_0\eta_\om(\bar{f}_0)/m_N^2
-U_{\Delta}(n_0)}{m_N-m_N^{*} (n_0)}\,,\quad x_{\om\Delta}=x_{\rho\Delta}=1\,, \quad x_{\phi\Delta}=0.
\label{x-QCU}
\end{eqnarray}
Here we  continue to use the quark counting relation for $x_{\om\Delta}$ and $x_{\rho\Delta}$ and the  Iizuka-Zweig-Okubo  suppression of the $\phi$ meson coupling to not strange baryons~\cite{Okubo}.

Unfortunately, the value $U_\Delta(n_0)$ is poorly constrained by existing data. Results of various analyses are contradictive.
From the analysis of electromagnetic excitations of $\Delta$s within a relativistic quantum-hadrodynamic scheme reference~\cite{Wehrberger1} concluded  that $0\lsim x_{\sigma\Delta}-x_{\om\Delta}\lsim 0.2$. Reference~\cite{Jin}  using the QCD sum rule  estimated  the coupling of the $\Delta$ to the $\om$ field to be half of the strength estimated from the quark counting, $x_{\om\Delta}\simeq 0.4$--0.5, whereas the coupling to the scalar field was estimated as $x_{\sigma \Delta}\simeq 1.3$. Calculations~\cite{Kosov} within the standard non-linear Walecka model showed that with such coupling parameters the ISM at $n=n_0$ would be metastable since there appears a second and much deeper minimum in the energy at the density $n\sim 3\,n_0$. These coupling parameters correspond to the potential $U_\Delta(n_0)$, being 3--5 times deeper than the nucleon potential.
A possibility for a large value of the $U_\Delta$ potential was advocated in~\cite{Connell},  where it was demonstrated that the electron--nucleus scattering can be described with
$U_\Delta (n_0)\simeq -115$\,MeV if the momentum dependence of the $\Delta$-nucleus potential is included. Following the relation (\ref{x-QCU}) the variation of the $\Delta$ potential in the interval $-150\,{\rm MeV}\le U_\Delta(n_0)\le -50\,{\rm MeV}$  corresponds to the variation $1.49(1.34)\ge x_{\sigma\Delta}\ge 0.94$ (0.94) for the KVORcut03(MKVOR) models (for $x_{\omega\Delta}=1$). Note that, if  we assume the same mass-scaling for $\Delta$s and nucleons, $\Phi_\Delta =\Phi_N$, that corresponds to $x_{\sigma \Delta}=1.32$,  we obtain $U_\Delta (n_0)\simeq -119$\,MeV for KVORcut03 and $U_\Delta (n_0)\simeq -146$\,MeV for MKVOR models.

However, it seems us rather unrealistic, if $\Delta$ baryons having similar internal quark structure as the nucleons had  feel
a much different potential. The same argumentation was used in~\cite{Migdal:1990vm,Voskresensky:1993ud} where the authors utilized $U_\Delta (n)\simeq U_N (n)$ with the nucleon potential $U_N (n_0)\simeq -(50\mbox{--}60)$\,MeV.

The coupling of the $\Delta$ baryon to the $\sigma$ field can be estimated, if one applies the chiral symmetry constraints to the $\pi\Delta$ scattering amplitude. A contribution to energy-independent isospin-symmetrical part of the pion-baryon scattering amplitude can be described from one side by the pion-baryon sigma-term and from the other side by the exchange of the $\sigma$ meson
\begin{eqnarray}
\frac{g_{\sigma B}g_{\sigma\pi\pi}}{m_\sigma^2} \approx \frac12\frac{\Sigma_{\pi B}}{f_\pi^2m_\pi}\,,
\label{gsbb-sigterm}
\end{eqnarray}
where  $\Sigma_{\pi B}$ is the pion-baryon sigma-term, $f_\pi$ is the pion decay constant, $m_\pi$ is the pion mass, and $g_{\sigma\pi\pi}$ is the $\sigma\pi\pi$ coupling constant. The similar relation was used in~\cite{BLRT94} [see Eq.~(23) there] for the kaon-nucleon scattering.
From the relation (\ref{gsbb-sigterm}) we estimate the coupling parameter
\begin{eqnarray}
x_{
\sigma\Delta} \approx {\Sigma_{\pi \Delta}}/{\Sigma_{\pi N}}.
\label{xs-sigt}
\end{eqnarray}
The sigma-terms are evaluated  in the quark model~\cite{Lubov} as
\begin{eqnarray}
\Sigma_{\pi N}= 43.3\pm 4.4\,{\rm MeV}\,,\quad
\Sigma_{\pi \Delta} =32\pm 3\,{\rm MeV}\,.
\label{sigterm}
\end{eqnarray}
Calculations in the framework of the chiral perturbation theory~\cite{Cavalcante} give  similar results
$\Sigma_{\pi N}= 45.8\,{\rm MeV}$, and $\Sigma_{\pi \Delta} =32.1\,{\rm MeV}$.
Equation~(\ref{xs-sigt}) with the parameters (\ref{sigterm}) yields the interval for the $x_{\sigma\Delta}$ values,  $ 0.90\gsim x_{\sigma\Delta} \gsim 0.61$\,. The latter interval corresponds to a shallow attractive or even repulsive $\Delta$ potential $-43(-40)\,{\rm MeV}\lsim U_\Delta(n_0)\lsim +10(+33)\,{\rm MeV}$ for our KVOR(MKVOR) models, provided we take $x_{\omega\Delta}=1$.

Studying the electron-nucleus scattering data Ref.~\cite{Koch1} introduced a density-dependent average binding potential  $U_\Delta (n)\simeq - 55\, n(r)/n_0$\,MeV. Reference~\cite{Nakamura} supported this estimate from the analysis of neutrino-induced pion production on carbon. On the other hand, from the study of the pion-nucleus scattering data Ref.~\cite{Horikawa} concluded  that the real part of the $\Delta$-nucleus potential is as shallow as $-30$\,MeV. Similar estimation is suggested in Ref.~\cite{EricsonWeise}. Since pions interact mainly close to the nucleus surface, larger values of the potential are expected at $n_0$, so for the linear density dependence one may expect that $U_{\Delta}(n_0)\sim U_N (n_0)$, in agreement with estimates~\cite{Migdal:1990vm,Koch1}.  Following analyses of electron-nucleus~\cite{Koch,Connell,Wehrberger} and pion-nucleus~\cite{Horikawa,Nakamura} scattering and photoabsorption~\cite{Alberico} the authors in~\cite{Drago2014} estimated a range of uncertainty for the $\Delta$ potential as  $-30\,{\rm MeV}+ U_N (n_0)< U_{\Delta}(n_0)<U_N(n_0)$ that with  $U_N (n_0)\simeq -(50\mbox{--}60)$\,MeV leads to the constraint $-90$\,MeV$< U_{\Delta}(n_0)<-50$\,MeV. The authors~\cite{Song,Ferini,Cozma,Guo} studying threshold conditions for pion and $\Delta$ productions in heavy-ion collisions arrived at inequality $U_N(n_0)<U_{\Delta}(n_0)<\frac{2}{3}U_N(n_0)$ that leads to inequality $-60\,{\rm MeV} < U_{\Delta}(n_0)<-40$\,MeV.
The most involved calculation in~\cite{Riek:2008uw} basing on a self-consistent and covariant many-body approach for the pion and $\Delta$ isobar propagation in ISM, from the study of  the photoproduction off nuclei adjusted the set of Migdal parameters and predicted $U_{\Delta}(n_0)=-50$\,MeV.

Below we will use the value $-50$\,MeV as a most realistic estimate of the $\Delta$ potential. We shall see that in this case effects of $\Delta$s within our models of EoS prove to be not so strong. To test the limits of the models  we also
allow for an enhancement of the $U_{\Delta}(n_0)$ attraction varying it in the interval $-(-150\mbox{--}100)\,{\rm MeV}\le U_\Delta(n_0)\le -50\,{\rm MeV}$.

For $\xi_{\sigma \Delta}=0$ at $n>n_{c,\Delta}$, that corresponds to  $m^{*}_\Delta =m_\Delta$ for $n>n_{c,\Delta}$,  the $\Delta$ baryons do not appear in any of the models considered below. Therefore, studying possible $\Delta$ effects on the EoS  we refuse this possibility and exploit a more realistic choice of $\xi_{\sigma \Delta}=1$ throughout the text.
In Sect.~\ref{Numerical} we use the traditional choice for the $\Delta$ coupling constants, $x_{\omega\Delta}=x_{\rho\Delta}=1$, and then in Sect.~\ref{sec:variation} we allow for their variation.

\section{KVORcut03, MKVOR and MKVOR* models }\label{sec:eos}

We focus now on two models KVORcut03 and MKVOR proposed in~\cite{Maslov:2015msa,Maslov:2015wba}, which proved to satisfy well many constraints on the hadronic EoS, and we extend them now including $\Delta$ baryons.
These two models utilize so called cut-mechanisms  of slowing down  the growth of $f$ field after it reaches some value with the density increase. The cut mechanism allows to stiffen the EoS, as was recently demonstrated in~\cite{Maslov:cut}.
In the KVORcut03 model it is achieved by a sharp variation of the $\eta_\om (f)$ scaling function, whereas in the MKVOR model a sharp variation is included in the $\eta_\rho (f)$ scaling function.
The latter is done to keep the EoS not too stiff in ISM to fulfill the flow constraint from heavy-ion collisions~\cite{Danielewicz:2002pu} and to make the EoS as stiff as possible in BEM to safely satisfy the constraint on the maximum mass of a compact star. The $\rho$ field is coupled to the isospin density that makes the $f$-saturation mechanism very sensitive to the composition of the BEM. As we shall see, the incorporation of $\Delta$ baryons  leads in the MKVOR model (now labeled as MKVOR$\Delta$ model)  to a problem that the nucleon effective mass in ISM drops to zero at some density (e.g., at $n\sim 6\,n_0$ for $U_\Delta (n_0)\sim -50$\,MeV) and for higher densities the description in terms of hadronic degrees of freedom becomes invalid. To prolong the hadronic description in ISM for higher densities we propose below a minimal modification of the MKVOR model (labeled  MKVOR*), which prevents the effective nucleon mass from vanishing at any density.
\begin{table}[b]
\caption{Coefficients of the energy expansion (\ref{Eexpans}) near $n_0$ for KVORcut03 and MKVOR models.}
\begin{center}
\begin{tabular}{ccccccccc}
\hline
\raisebox{-.23cm}[0cm][0cm]{EoS}& $\mathcal{E}_0$ & $n_0$ & $K$ & $m_N^*(n_0)$ &$J$ & $L$ &$K'$ & $K_{\rm sym}$
\\ \cline{2-9}
&           [MeV] & [fm$^{-3}$] & [MeV] & $[m_N]$  & [MeV] & [MeV] &[MeV] & [MeV]
\\ \hline
KVORcut03 & $- 16$ & 0.16 & 275 &  0.805 &  32 & 71 &  422& -86\\
MKVOR     & $- 16$ & 0.16 & 240 &  0.730 &  30 & 41 &  557 & -158\\
\hline
\end{tabular}
\end{center}
\label{tab:sat-param}
\end{table}


The properties of our model at the nuclear saturation density $n_0$ are illustrated
in Table~\ref{tab:sat-param}, where we collect coefficients of the expansion of the nucleon binding energy per nucleon near $n_0$ for KVORcut03 and MKVOR models,
\begin{align}
& \mathcal{E} = \mathcal{E}_0 + \frac{1}{2}K\epsilon^2
-\frac{1}{6}K'\epsilon^3 +\beta^2\widetilde{\mathcal{E}}_{\rm sym}(n) +
O(\beta^4, \epsilon^4)\,,
\nonumber\\
& \widetilde{\mathcal{E}}_{\rm sym}(n)=J + L\epsilon +\frac{K_{\rm sym}}{2}\epsilon^2+\dots\,,
\label{Eexpans}
\end{align}
in terms of small $\epsilon=(n-n_0)/3n_0$ and $\beta=(n_n-n_p)/n$ parameters. The parameters for the MKVOR* and MKVOR models are identical.

\subsection{KVOR and KVORcut models}

\begin{figure}
\centering
\includegraphics[width=14cm]{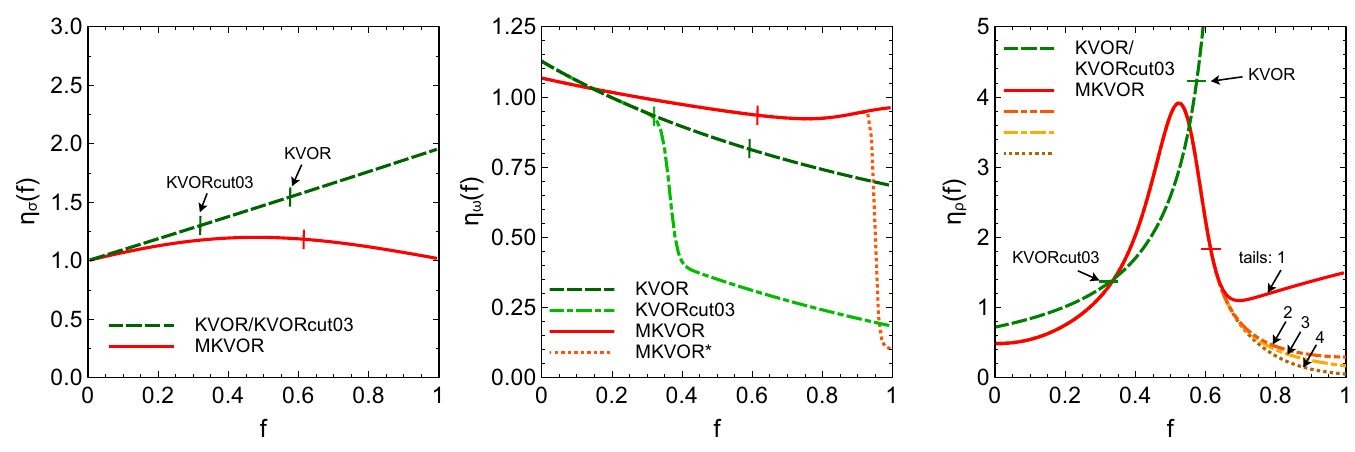}
\caption{Scaling functions $\eta_\sigma$ (left panel), $\eta_\om$ (middle panel), and $\eta_\rho$ (right panel) as functions of the scalar field $f$ for the KVOR, KVORcut03, MKVOR and MKVOR*  models. For the $\eta_\rho (f)$  we show also  variations of the function defined in~(\ref{zetaf}) with parameters (\ref{tail123}).
Vertical and horizontal bars indicate the maximum values of $f$ ($f_{\rm lim}$)  reachable at densities available in NSs.}
\label{Fig-1-new}
\end{figure}

Now we introduce the scaling functions.  First, we remind the choice for scaling functions in the KVOR model~\cite{Kolomeitsev:2004ff}:
\begin{eqnarray}
&& \eta^{\rm KVOR}_\sigma = 1 + 2 \frac{C_\sigma^2}{ f^{2}}\,  \big(\frac{b}{3} f^3 + \frac{c}{4} f^4\big)
\,,\quad
\eta^{\rm KVOR}_\omega = \Big[\frac{1 + z \bar{f}_0}{1 + z f}\Big]^\alpha\,,\quad \bar{f}_0=f(n_0)\,,
\nonumber\\
&& \eta^{\rm KVOR}_\rho = \Big[1 + 4\,\frac{C_\om^2}{C_\rho^2}\,(1-[\eta^{\rm KVOR}_\om (f)]^{-1})\Big]^{-1}\,.
\label{eta-KVOR}
\end{eqnarray}
The scaling functions (\ref{eta-KVOR}) are plotted in Fig.~\ref{Fig-1-new}.
The $\eta^{\rm KVOR}_\sigma$ function is just a  reparametrization of the $\sigma$ self-interaction potential $U(f)$ proposed by Boguta and Bodmer~\cite{Boguta77} in terms of the scaling function. The function $\eta^{\rm KVOR}_\omega$ is chosen to be a decreasing function of $f$
smaller than 1 for $f>f_0$, that leads to an increase of the $\omega$ meson contribution to the energy density and to a stiffening of the EoS.
The choice of $\eta^{\rm KVOR}_\rho$ is made to guarantee a  monotonous decrease of the effective nucleon mass with a density increase in the BEM for densities relevant for NSs. Such a $m^*(n)$ decrease is in a line with ideas of the partial restoration of the chiral symmetry and Brown-Rho scaling. An increase of $\eta^{\rm KVOR}_\rho$ with increase of $f$   allows to suppress the symmetry energy and  the proton fraction in the NS  for $n>n_0$, helping to fulfill the DU constraint on the efficiency of the NS cooling, cf.~\cite{Blaschke:2004vq,Kolomeitsev:2004ff,Grigorian:2005fn,Klahn:2006ir,Grigorian:2016leu}.

For the KVORcut models the scaling functions were chosen in~\cite{Maslov:2015wba} in the following form
\begin{eqnarray}
&&\eta^{\rm KVORcut}_\sigma(f) = \eta^{\rm KVOR}_\sigma \,,\quad
\eta^{\rm KVORcut}_\omega(f) = \eta^{\rm KVOR}_\omega + a_\omega \theta_{b_\om}(f-f_\om)
\,,
\nonumber\\
&&\eta^{\rm KVORcut}_\rho (f) =\eta^{\rm KVOR}_\rho\,.\quad
\label{eta-KVORcut}
\end{eqnarray}
We introduced here the switch functions
\begin{eqnarray}
\theta_y(x)=\frac12 \big[1+ \tanh(y x)\big]
\end{eqnarray}
with the limits $\theta_y(-\infty)=0$ and $\theta_y(+\infty)=1$. In the limit $y\to +\infty$ this function turns into the Heaviside step function, $\theta_y(x)\to (1+{\rm sign}(x))/2$. For the model KVORcut03, parameters determining  the scaling functions and the EoS, see Eq. (\ref{En}), are collected in Table~\ref{tab:param-KVORcut03}, $\bar{f}_0=f(n_0)$.

\begin{table}
	\caption{Parameters of the  KVORcut03 model.}
	\begin{center}
		\begin{tabular}{lccc ccc ccc }
			\hline\hline
$C_\sigma^2$ & $C_\om^2$ & $C_\rho^2$ & $b \cdot 10^3$ & $c \cdot 10^3$  & $\alpha$ & $z$& $a_\om$  &$b_\om$ & $f_\om$ \\   
\hline
179.56& 87.600& 100.64& 7.7354& 0.34462& 1 & $-$0.5& 0.11 & 46.78 & 0.365 \\
			\hline \hline
		\end{tabular}
	\end{center}\label{tab:param-KVORcut03}
\end{table}

\begin{figure}
\centering
\includegraphics[width=5cm]{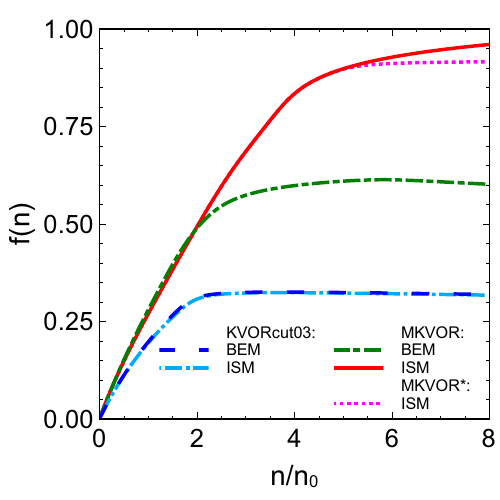}
\caption{
 Scalar field $f$ as a function of the nucleon density $n$ in the ISM and BEM for KVORcut03, MKVOR, and MKVOR* models. Note that in BEM the functions $f(n)$ for MKVOR and MKVOR* models are identical.}
\label{fig:eta_r}
\end{figure}

Functions $\eta_\sigma(f)$, $\eta_\omega(f)$, $\eta_\rho (f)$ for models which we consider in the given paper are presented in Fig.~\ref{Fig-1-new}. Vertical and horizontal bars indicate the maximum values of $f$  reachable in NSs  for the EoSs under consideration. For these EoSs they correspond to central densities for stars with $M=M_{\rm max}$.
For the models KVOR and KVORcut03 the functions $\eta_\rho (f)$, $\eta_\sigma(f)$ are smooth functions of $f$ and the cut-procedure is applied to the $\eta_\omega(f)$, which decreases rapidly in the interval $0.3 < f < 0.4$\,.  The field variable $f$ proves to be restricted from above by the value $f_{\rm lim}$ (being slightly above 0.3) and very weakly depends on the isospin composition of the matter.   With $\eta_\sigma (f)$, $\eta_\rho (f)$ and $\eta_\sigma(f)$ functions under consideration there is a single solution $f(n)$.
The functions $f(n)$, being solutions of Eq.~(\ref{eq_fn}) for ISM and BEM, are shown in Fig.~\ref{fig:eta_r} (left panel). For the KVORcut03 model in  both cases $f(n)$ grows from zero at $n=0$ to the value $\simeq 0.3$ at $n\simeq 2n_0$,  and with a further increase of the density the growth is terminated at the limiting value $f_{\rm lim}$, which is slightly above 0.3 both in ISM and BEM.

We checked that in the KVOR and KVORcut, also in KVOR- and KVORcut- based models, when hyperons and $\Delta$ baryons are included, Eq.~(\ref{eq_fn}) for $f$ has only one solution for any density and equilibrium isospin composition.

\subsection{MKVOR and MKVOR* models}

The model MKVOR  proposed in~\cite{Maslov:2015msa,Maslov:2015wba} is characterized by the following scaling functions:
\begin{align}
\eta^{\rm MKVOR}_\sigma(f) &= \Big[1 - \frac{2}{3} C_\sigma^2 b f -
\frac{1}{2} C_\sigma^2 \Big(c -
\frac{8}{9} C_\sigma^2 b^2\Big) f^2 + \frac{1}{3} d f^3\Big]^{-1} \,,
\nonumber\\
\eta^{\rm MKVOR}_\omega(f) &= \eta^{\rm KVORcut}_\omega(f)\,,
\label{eta-MKVOR}\\
\eta^{\rm MKVOR}_\rho(f) &= a_\rho^{(0)} + a_\rho^{(1)} f +
\frac{a_\rho^{(2)} f^2}{1 + a_\rho^{(3)} f^2}  +
\beta \exp\big(- \Gamma(f)(f - f_\rho)^2 \big)\,,
\nonumber\\
\Gamma(f) &= {\gamma }
\Big[{1 + \frac{d_\rho (f-\bar{f}_0)}{1 + e_\rho (f-\bar{f}_0)^2}
}\Big]^{-1}\,,
\nonumber
\end{align}
with the parameters listed in Table~\ref{tab:param-MKVOR}.

The scaling functions $\eta_\sigma (f)$, $\eta_\om (f)$ and $\eta_\rho (f)$ for the MKVOR model are shown in Fig.~\ref{Fig-1-new} (see also Fig.~4 in~\cite{Maslov:2015wba}). Vertical and horizontal bars indicate the maximum values of $f$ reachable in NSs with the maximum masses. In the MKVOR model the ``cut'' mechanism limiting the growth of the $f$ field with a density increase is not operative in ISM, since $\eta_\sigma (f)$, $\eta_\om (f)$ are chosen as smooth functions of $f$. The strong variation of the scaling with $f$ is implemented in this model in the $\rho$-meson sector  (in the $\eta_\rho (f)$ function). The $\rho$-meson term does not contribute in ISM. Oppositely, in the BEM the magnitude of the scalar field $f(n)$ becomes limited from above. This mechanism allows us to push up the maximum NS mass and simultaneously satisfy the constraint deduced from the analysis of the particle flows in heavy-ion collisions. The $\eta_\rho (f)$ determined by  Eq.~(\ref{eta-MKVOR}) with parameters from Table~\ref{tab:param-MKVOR} is indicated in Fig.~\ref{Fig-1-new} by ``tail 1".

\begin{table}
	\caption{Parameters of the MKVOR model.}
	\begin{center}
		\begin{tabular}{lccc ccc ccc}
			\hline\hline
$C_\sigma^2$ & $C_\om^2$ & $C_\rho^2$ & $b \cdot 10^3$ & $c \cdot 10^3$ &  $d$ &$\alpha$ & $z$& $a_\om$ & $b_\om$ \\
\hline
234.15&134.88&81.842&4.6750&$-$2.9742&$-$0.5&0.4&0.65& 0.11 &  7.1\\
\hline
$f_\om$	& $\beta$& $\gamma$ & $f_\rho$ & $a_\rho^{(0)}$& $a_\rho^{(1)}$ & $a_\rho^{(2)}$ & $a_\rho^{(3)}$ & $d_\rho$ & $e_\rho$ \\
			\hline
0.9 & 3.11 & 28.4 & 0.522 & 0.448 & $-$0.614 & 3   & 0.8 & $-$4 & 6 \\
			\hline \hline
		\end{tabular}
	\end{center}\label{tab:param-MKVOR}
\end{table}

\begin{figure}
\centering
\includegraphics[width=14cm]{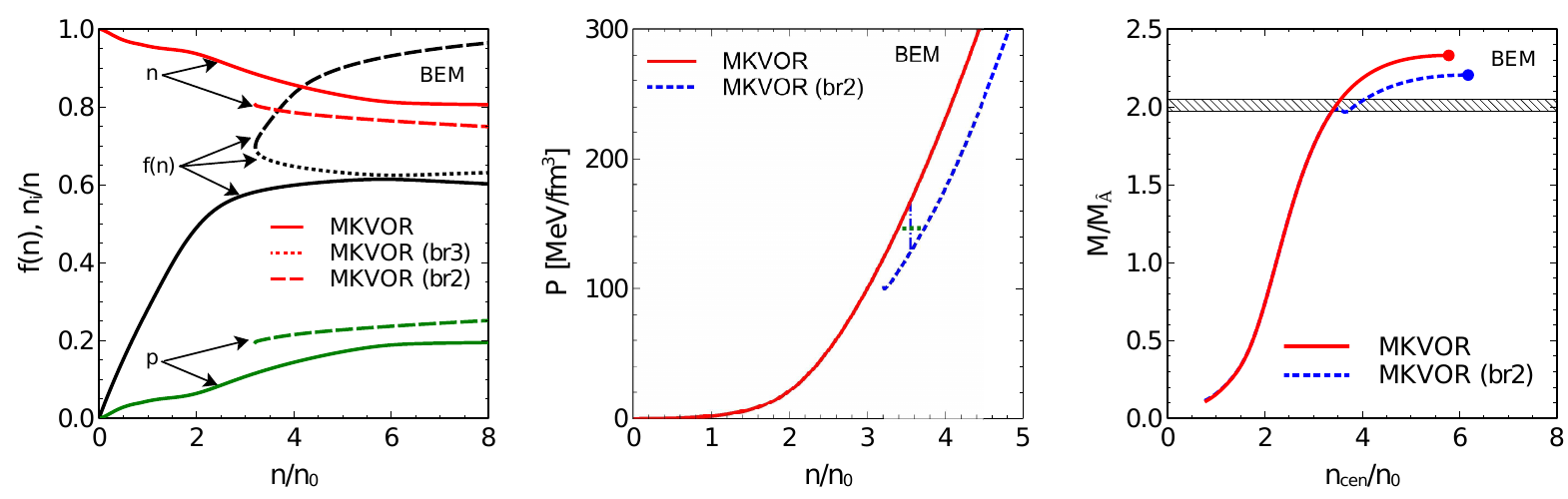}
\caption{ {\em Left panel:} Nucleon concentrations and magnitude of the scalar field, $f(n)$, as functions of the nucleon density in the BEM for the MKVOR model. For $n>3.21 n_0$ besides the original branch 1 (labeled as MKVOR) with the limit $\lim_{n\to 0}f(n)= 0$, there appear extra two branches 2,3 labeled as MKVOR(br2) and  MKVOR(br3). Branches 1,2 correspond to local minima in $E(f)$, whereas branch 3, to a local maximum. Nucleon concentrations are shown for branches 1 and 2 only. {\em Middle panel:} pressure $P(n)$ for branches 1 and 2. Vertical line indicates points of equal energies, horizontal line is the MC line.  {\em Right panel:} The NS mass as a function of the central density for the  branch 1 (MKVOR) and for the EoS with a first-order phase transition from the MKVOR branch to the MKVOR(br2) branch. }
\label{MKVORbranches}
\end{figure}

A general comment concerning scaling functions is in order. The $\eta_\om (f)$ and $\eta_\rho (f)$ functions  for the KVOR model were chosen originally in~\cite{Kolomeitsev:2004ff}  in a rather simple form (\ref{eta-KVOR}) following the pragmatic reasons. For such a choice of the scaling functions in the KVOR and KVORcut-based models there exists always only one solution of Eq.~(\ref{eq_fn}) for any $n$. In the MKVOR model  a  more complicated  $f$-dependence of the scaling functions is chosen to  satisfy the known experimental constraints, especially to better fulfill simultaneously the flow and maximum compact star mass constraints. In~\cite{Maslov:2015msa,Maslov:2015wba} we used the solution $f(n)$, which starts from the origin  $f=0$, $n=0$.
However, for the original choice of the $\eta_\rho (f)$ function (shown in Fig.~\ref{Fig-1-new} by the line labeled with ``tail 1'')  besides the solution starting at  the origin (branch 1) there appear two new solutions (branches 2 and 3) for densities  $n>3.21 n_0$. All these branches of solutions for $f(n)$ in BEM are depicted on the left panel of Fig.~\ref{MKVORbranches}. Branches 1, 2, and 3 are determined as zeros of the function $D(f,n)=\frac{\partial E(f,n)}{\partial f}$.
For branches 1 and 2 we find $(\frac{\partial D(f)}{\partial f})_{f_{1,2}}>0$
and hence branches 1 and 2 correspond to minima of the energy-density functional $E(f)$. Oppositely for the branch 3 we have $(\frac{\partial D(f)}{\partial f})_{f_3}<0$ and therefore this branch is related to a maximum of $E(f)$. Thus, the branch 3 can be disregarded. On the left panel of Fig. \ref{MKVORbranches} we also show  the neutron and proton concentrations for  branches 1 (labeled as MKVOR) and 2 (labeled as MKVOR(br2)). On the middle panel of Fig. \ref{MKVORbranches} we show the pressure of the BEM as a function of density, $P(n)$, for branches 1 and 2. At the densities $n<n_1^{\rm MC}$ the system follows the branch 1 (line MKVOR). Transition from  the branch 1 to the branch 2 is a first-order phase transition. Within the density range $n_1^{\rm MC}<n<n_2^{\rm MC}$ the pressure and baryon chemical potential  follow the Maxwell construction (MC) line determined by equations $P(n)=P(n_1^{\rm MC})=P(n_2^{\rm MC})$ and $\mu_B (n) =\mu_B (n_1^{\rm MC}) =\mu_B (n_2^{\rm MC})$.\footnote{Here, we disregard a possibility of a mixed pasta phase following an observation of \cite{VYT} that with taking into account of finite size effects the description of the pasta phase might be  close to description given by the MC.} For $n>n_2^{\rm MC}$ the system follows the branch 2 (line MKVOR(br2)). The vertical line indicates points of equal energy.
On the right panel of Fig.~\ref{MKVORbranches} we show the NS mass as a function of the central density. We see that the NS configurations constructed with $f(n)$ taken along the branch 1 (solid line) would lead to  a higher NS mass at fixed central density than those constructed with the transition from the branch 1 to the branch 2 (dashed line). Thus the first-order phase transition  from the branch 1 to the branch 2  is indeed energetically favorable in the given model. Note that three-branch solutions appear also, when one considers the ordinary  RMF models in ISM at high temperature, see Fig. 3 in  \cite{Glendenning87}.

Working in the framework of the purely hadronic model we see no weighty reason for a phase transition to occur at $n$ of the order of several $n_0$ with  a jump in the scalar-field magnitude. Therefore, we will avoid this possibility in the given paper, although a further study of such a transition  might be of interest, if considered as a simplified model for a first-order hadron-quark phase transition.

In~\cite{Maslov:2015msa,Maslov:2015wba},  we considered only the solution with $f$ corresponding to branch 1. Other branches correspond to the values of $f(n)$ larger than $f_{\rm lim}$, where $f_{\rm lim}$ is the maximum value on branch 1 reachable in the BEM in the center of the NS with $M=M_{\rm max}$.
Therefore, additional unwanted solutions can be eliminated in the MKVOR and MKVOR-based models by  an appropriate variation of the $\eta_\rho$ function for $f>f_{\rm lim}$.
To demonstrate this we propose a modification of the $\eta_\rho$ scaling function
\begin{align}
&\eta_\rho^{\rm MKVOR}(f) \to \left\{
\begin{array}{lc}
\eta_\rho^{\rm MKVOR}(f)\,, & f\le f_\rho^*\\
1/[a_0
+ a_1 z
+ a_2 z^2
+ a_3 z^3
+ a_4 z^4]
   \,, & f >f_\rho^*
\end{array}
\right.\,,
\label{zetaf}\\
&\qquad\qquad\qquad  z=f/f_\rho^*-1\,, \quad f^*_\rho = 0.64 \,,
\nonumber
 \end{align}
where  we change its ``tail'' for $f>f_\rho^* >f_{\rm lim}$.
Parameters $a_{0}$, $a_{1}$, and $a_{2}$ follow from the continuity of the function and its first two derivatives in the point $f=f_\rho^*$:
\begin{eqnarray}
& a_0 = \eta^{-1}_\rho(f^*_\rho), \quad
a_1 = -f^*_\rho\,\eta_\rho'(f^*_\rho)\,a_0^2, \quad
a_2 =  a_1^2/ a_0 - a_0^2 \eta_\rho''(f^*_\rho)\, f^{*2}_\rho /2.
\label{tail-par1}
\end{eqnarray}
Here we skip the superscript MKVOR on $\eta_\rho$ for the sake of brevity.
Other parameters $a_3$ and $a_4$ control the slope of the tail of the scaling function.
So, together with the original parametrization (\ref{eta-MKVOR}), which we now label ``tail 1'',
we consider several other choices:
\begin{align}
&\mbox{tail 2}: a_3=-10\,,\quad a_4=0\,;
\nonumber\\
&\mbox{tail 3}: a_3=0\,,\phantom{-1}\quad a_4=0\,;
\label{tail123}\\
&\mbox{tail 4}: a_3=0\,,\phantom{-1}\quad a_4=100\,.
\nonumber
\end{align}
From now on, under the MKVOR model we will understand the model with $\eta_\rho$ having appropriate continuation for $f>f_{\rm lim}$ which removes multiple solutions, e.g. with one of tails 2, 3, or 4 shown in Fig.~\ref{Fig-1-new}. We have verified that for the choices (\ref{tail123}) the unwanted solutions with large values of $f$ are absent in all MKVOR-based models, which we studied  previously in ~\cite{Maslov:2015msa,Maslov:2015wba} (without and with hyperons) and consider below (without and with hyperons and $\Delta$s).
For $f<f_{\rm lim}<f_\rho^*$, the  $\eta_\rho(f)$ function coincides exactly with that for the originally introduced scaling function, see  Fig.~\ref{Fig-1-new}.

Below we will see that in the presence of $\Delta$ baryons, i.e., within the MKVOR$\Delta$ model, the effective nucleon mass vanishes at some density in the ISM. To cure this problem within our hadronic model we will introduce additional cut-mechanism in the $\om$ sector, keeping the same $\eta_\sigma(f)$ and $\eta_\rho(f)$ as in MKVOR model, the latter function with the tail modification (\ref{zetaf}) serving for the uniqueness of the $f(n)$ solution in BEM. In such a modified MKVOR model, which we label as MKVOR*, we use
\begin{align}\label{etaMKVOR*}
& \eta^{\rm MKVOR*}_\omega(f)  = \eta_\omega^{\rm MKVOR}(f)
 \theta_{b_\om}(f_\om^*-f) + \frac{c_\om}{(f/f_\om^*)^{\alpha_\om}+1} \theta_{b_\om}(f-f_\om^*)\,,
\nonumber\\
&f_\om^*=0.95\,,\quad b_\om=100\,,\quad \alpha_\om=5.515\,,\quad c_\om=0.2299\,.
\end{align}
For $f<f_\om^*$ the scaling function $\eta^{\rm MKVOR*}_\om(f)$ fits that for the original MKVOR model. For $f>f_\om^*=0.95$, $\eta^{\rm MKVOR*}_\om(f)$ sharply decreases. Thereby, we limit the rapid growth of the scalar field $f(n)$ with a density increase not only in BEM, as it was in the original MKVOR model, but also in the ISM.


The functions $f(n)$ for the MKVOR and MKVOR* models in ISM and BEM are demonstrated in Fig.~\ref{fig:eta_r}. In the BEM the cut mechanism, implemented in the MKVOR model in the $\rho$ sector, fixes the magnitude of the scalar field at the level $f_{\rm lim}\approx 0.6$, and $m_N^{*} $ reaches the minimum value $\simeq 0.4m_N$ for $n \gsim 4 n_0$. Since the chosen cut value $f_\om^*=0.95$ is larger than  $f_{\rm lim}$, all results for the MKVOR-based models and the corresponding MKVOR*-based models coincide exactly in BEM. In the ISM the effective nucleon mass continuously decreases in MKVOR model with a density increase (for $n=8\,n_0$ it reaches $\simeq 0.05\,m_N$).  In the MKVOR* model the cut-mechanism is implemented in the $\omega$ sector and is operative in ISM.  With $f_\om^*=0.95$, for $n=8\,n_0$ we have $m^*_N\simeq 0.1\,m_N$. The saturation in $f(n)$ sets in only for $n\gsim 5\,n_0$  and for smaller densities the quantities $f(n)$ in the MKVOR and MKVOR* models follow the same curve in ISM. Due to that the nucleon and kaon flow constraints, which restrict  the allowed range for pressure in ISM in the density interval  $n_0< n\lsim 4.5\, n_0$, see~\cite{Maslov:2015msa,Maslov:2015wba}, are fulfilled in the MKVOR* model as well as in MKVOR one.

The following remark is in order. Unless we take into account finite-size effects, the effective meson masses and coupling constants enter the energy density functional only in $\eta_M $ combinations.  Thus, we can extract the $\chi_M (n)$ dependence only if we assume particular dependence $m^*_M(f(n))$, as the Brown-Rho scaling law (\ref{PhiN}) in our case. Varying the latter we would get different functions $\chi_M (n)$.

\section{Results of numerical calculations}\label{Numerical}

\subsection{KVORcut03-based models}

\begin{figure}[!h]
\centering
\includegraphics[width=5cm]{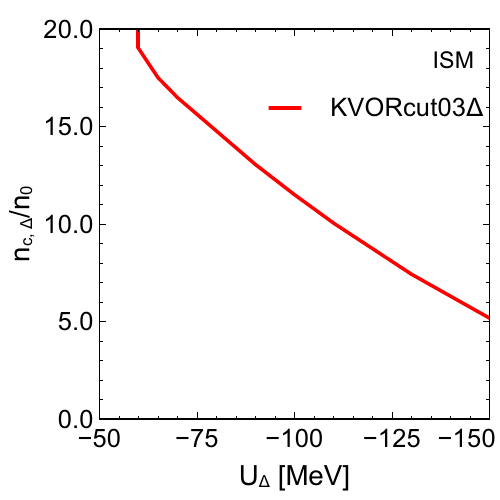}
\caption{Critical density for the appearance of $\Delta$ baryons, $n_{c,\Delta}$,
 as a function of the $\Delta$ potential $U_\Delta$ in the ISM
for the KVORcut03$\Delta$ model with the $\Delta$ parameter set (\ref{x-QCU}).
}
\label{fig:ncD-ISM-cut03}
\end{figure}

First we consider the influence of the presence of $\Delta$ baryons in ISM. In contrast to the standard non-linear Walecka models~\cite{Boguta1982,Kosov} the KVORcut03 model proves to be much less sensitive to the inclusion of $\Delta$ baryons.
For the parameter set (\ref{x-QCU}) the critical density for the  appearance of $\Delta$s in the KVORcut03$\Delta$ model is shown in Fig.~\ref{fig:ncD-ISM-cut03}. We see that for the realistic values of the potential\footnote{Shortening notation, below we will use $U_\Delta$ instead of $U_\Delta (n_0)$.} ($U_\Delta\gsim -60\,{\rm  MeV}$) the $\Delta$ baryons  do not appear in the ISM up to very high densities.
The reason for the robustness of the KVORcut03 model against  the $\Delta$ appearance is that the $f(n)$ stops to grow for densities $n\gsim 2\, n_0$ and has a smaller magnitude that would be in the non-linear Walecka models with the same $m_N^*(n_0)$, see Figs.~1--3 in~\cite{Maslov:2015wba}. This is a genuine feature of all ``cut''-models, which we have considered in~\cite{Maslov:2015wba}. As the result, the effective $\Delta$ mass remains rather large  that inhibits the growth of the $\Delta$ population.

\begin{figure}
\centering
\includegraphics[width = 14cm]{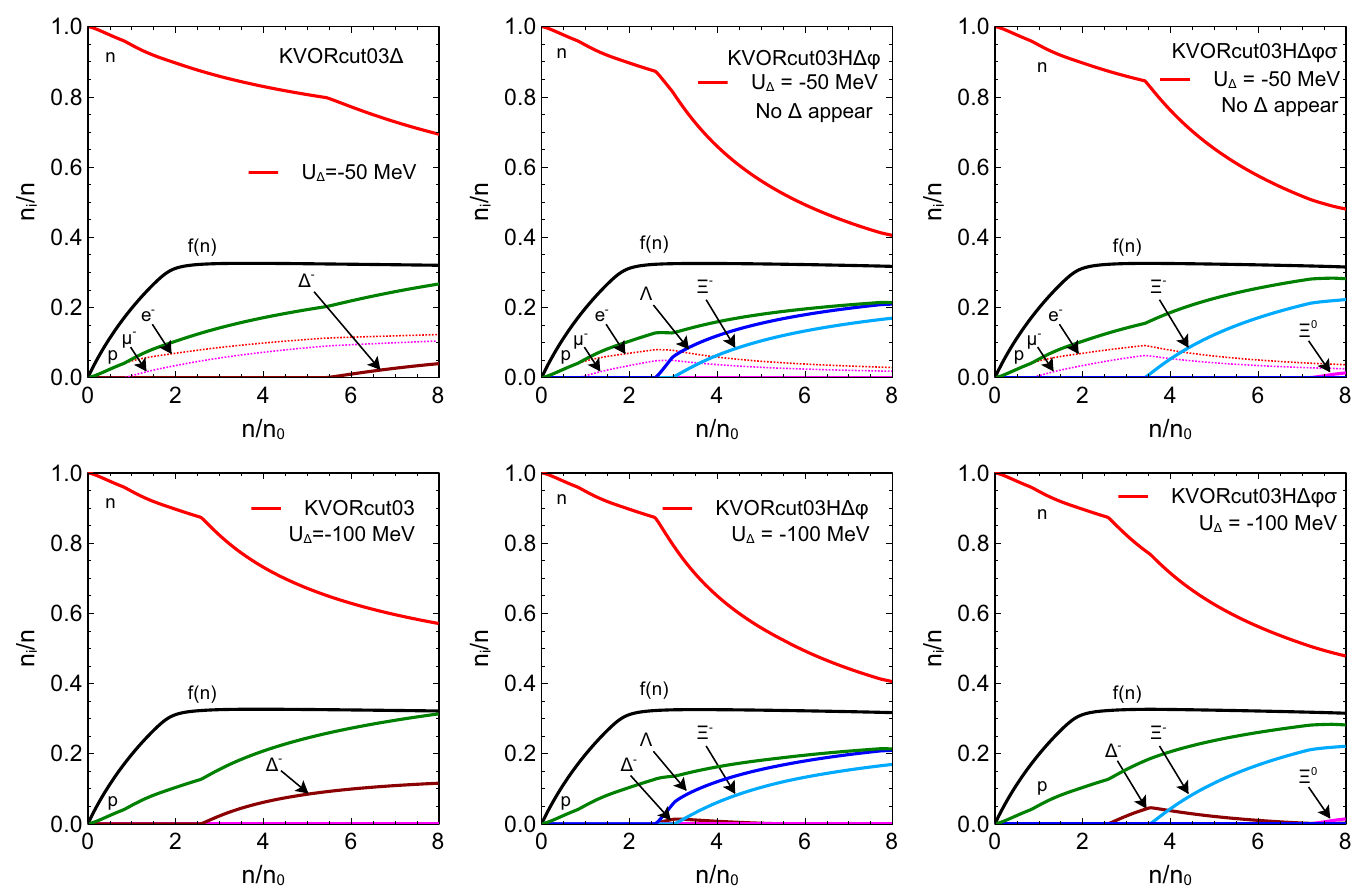}
\caption{Baryon concentrations and magnitude of the scalar field, $f(n)$, in the BEM for the KVORcut03$\Delta$, KVORcut03H$\Delta\phi$, and KVORcut03H$\Delta\phi\sigma$ models
for $\Delta$ potentials $U_{\Delta}=-50$\,MeV (upper row) and $U_{\Delta}=-100$\,MeV (lower row). The $\Delta$ parameters are taken as in Eq.~(\ref{x-QCU}).}
\label{fig:cut03-conc}
\end{figure}

In Fig.~\ref{fig:cut03-conc} we show the composition of BEM vs. the total baryon density for three different versions of KVORcut03 model:  with $\Delta$ baryons only (hyperons are artificially excluded) labeled KVORcut03$\Delta$ and with $\Delta$s and hyperons, incorporated according to the schemes (\ref{hyp-Hphi}) and (\ref{hyp-Hphisig}), labeled as KVORcut03H$\Delta\phi$ and KVORcut03H$\Delta\phi\sigma$, respectively.  In the KVORcut03$\Delta$ model
the $\Delta^-$ baryons appear in the BEM for the realistic value of the potential $U_\Delta=-50$\,MeV at densities $n>n_{{\rm c},\Delta^-} \simeq 5.4\, n_0$. Other $\Delta$ species ($\Delta^0$ and $\Delta^+$) do not appear up to maximum densities reachable in NS interiors. In the presence of hyperons, i.e., in the KVORcut03H$\Delta\phi$ and KVORcut03H$\Delta\phi\sigma$ models, $\Delta$ baryons do not appear. Similar inhibiting action of hyperons on the $\Delta$ population was noticed  also in~\cite{Drago2014}. For  the $\Delta$ potential of $-100$\,MeV, in all models  $\Delta^-$s  appear at approximately the same density, $n_{{\rm c},\Delta^-}\simeq 2.6\,n_0$.
  In the KVORcutH$\Delta\phi$ model $\Delta^-$  appear at the same critical density as $\Lambda$'s. In the  KVORcutH$\Delta\phi\sigma$ model $\Delta^-$s  appear before hyperons. In both cases in the presence of hyperons the $\Delta^-$ concentration remains tiny (does not exceed 5\%). Other $\Delta$ species ($\Delta^0$ and $\Delta^+$) do not appear in KVORcut03-based models in NSs.

\begin{figure}\centering
\includegraphics[width=10cm]{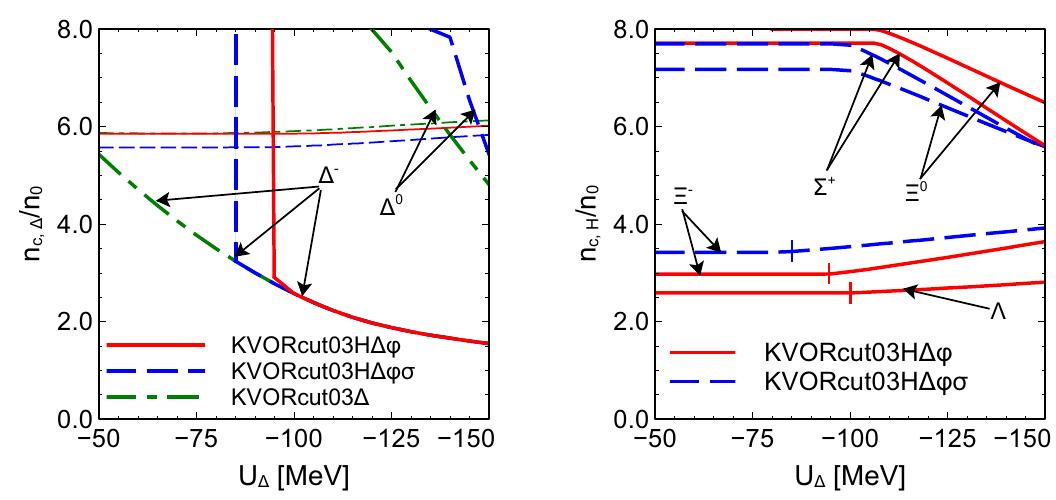}
\caption{Dependence  of the critical density for the $\Delta$  (left panel) and hyperon (right panel) appearance in BEM on the $\Delta$ potential for the KVORcut03$\Delta$, KVORcut03H$\Delta\phi$, and KVORcut03H$\Delta\phi\sigma$ models with the $\Delta$ parameters given by Eq.~(\ref{x-QCU}). Vertical  bars on the right panel indicate densities, at which  $n_{{\rm c},\Delta^-}$ coincides with the critical density of the corresponding hyperon species.}
\label{fig:cut03-Udep}
\end{figure}

In Fig.~\ref{fig:cut03-Udep} we show the dependence of critical densities for the appearance of $\Delta^-$ and $\Delta^0$ baryons (left panel) and those for hyperons (right panel) on the value of the $\Delta$ potential.
 Vertical bars on right panel indicate densities at which  $n_{{\rm c},\Delta^-}$ coincides with the critical density of the corresponding hyperon species.
In the KVORcut03$\Delta$ model the value  $n_{{\rm c},\Delta^-}$  monotonously decreases from $n_{{\rm c},\Delta^-}= 5.4 \, n_0$ for  $U_\Delta = -50$~MeV to $n_{{\rm c},\Delta^-}=  2.3 \, n_0$ for  $U_\Delta = -100$~MeV, and to  $n_{{\rm c},\Delta^-}= 1.6 \, n_0$ for $U_\Delta = -150$~MeV, the latter deep potential we consider as unrealistic.
In the KVORcut03H$\Delta\phi$ model at $U_\Delta > -95$~MeV and in the KVORcut03H$\Delta\phi\sigma$ models  at $U_\Delta > -85$~MeV $\Delta$s do not  appear at any relevant  densities. For $U_\Delta <-100$~MeV and for $U_\Delta <-85$~MeV models
KVORcut03H$\Delta\phi$ and KVORcut03H$\Delta\phi\sigma$, respectively, follow the same curve as KVORcut03$\Delta$. This happens because for the KVORcut03H$\Delta\phi$ model at $U_\Delta < -100$\,MeV (for the KVORcut03H$\Delta\phi\sigma$ model at $U_\Delta < -85$\,MeV) the critical density for $\Delta^-$ becomes smaller (see the right panel of Fig.~\ref{fig:cut03-Udep}) than the smallest among critical densities for hyperons and the latter ones do not inhibit the $\Delta$ population thereby.   On the right panel we also see that in the KVORcut03H$\Delta\phi$ model the hyperon species appear in the BEM with a growth of the density in the following order: first $\Lambda$s, then $\Xi^-$s after them $\Sigma^+$s and $\Xi^0$s as the latest ones. In the KVORcut03H$\Delta\phi\sigma$ model the order changes: there are no $\Lambda$s,
$\Xi^-$s appear first, then $\Xi^0$ and then $\Sigma^+$s.

\begin{figure}
\centering
\includegraphics[width=10cm]{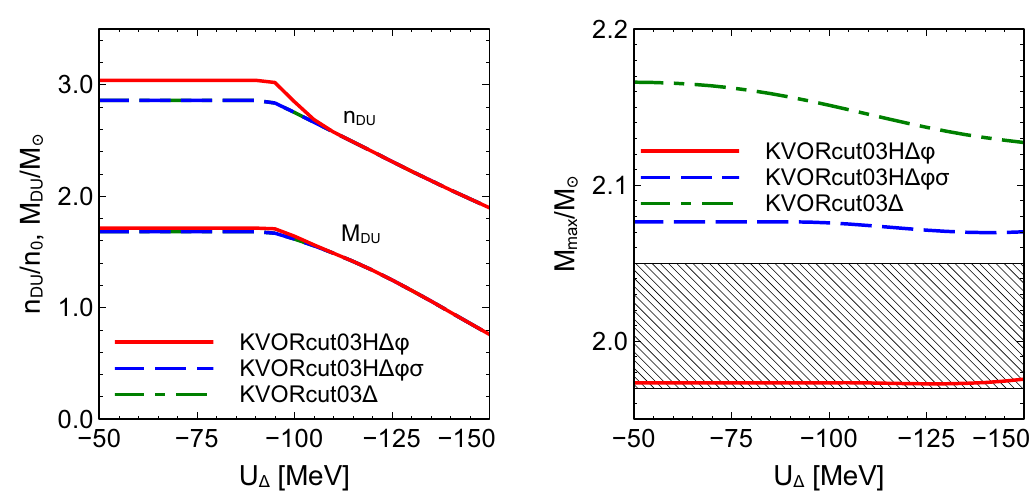}
\caption{
Critical density and critical NS mass for the DU reactions on nucleons  (left panel)  and the maximum NS mass (right panel) as functions of the $\Delta$ potential for the KVORcut03$\Delta$, KVORcut03H$\Delta\phi$, and KVORcut03H$\Delta\phi\sigma$ models with the $\Delta$ parameters given by Eq.~(\ref{x-QCU}). On the left panel  the curves for  the KVORcut03$\Delta$ and KVORcut03H$\Delta\phi\sigma$ models coincide.
The horizontal band on the right panel shows the uncertainty range for the measured mass of PSR J0348+0432
($2.01\pm 0.04\,M_\odot$).
  }
\label{fig:cut03-Udep-1}
\end{figure}

As demonstrated in Ref.~\cite{Maslov:2015wba}, the  critical density for DU processes on nucleons for the KVORcut03 model is $2.85\,n_0$ with the corresponding star mass $1.68\,M_\odot$. The critical star masses for the DU reactions on hyperons in KVOR\-cut03\-H$\phi$ and  KVORcut03H$\phi\sigma$ models are $1.51\,M_\odot$  and $1.91\,M_\odot$, respectively. So, these models satisfy both the ``weak" ($M>1.35 M_{\odot}$) and ``strong" ($M>1.5 M_{\odot}$) DU constraints introduced in~\cite{Kolomeitsev:2004ff,Klahn:2006ir}.
The presence of $\Delta$ baryons would shift the critical densities for the appearance of hyperons and, therewith, the critical densities for the processes involving them, e.g.,
$H\to N+l^-+\bar{\nu}_l$ and $\Delta^-\to\Lambda + e +\bar{\nu}_e$, to even higher values.
As pointed out in~\cite{Prakash-DU} the DU processes on $\Delta^-$ ($\Delta^-\to n+ l^- +\bar{\nu}_l$) are forbidden, if the DU processes on nucleons are forbidden because $n_{\Delta^-}<n_p$.
Therefore, to understand, if our model with $\Delta$ baryons satisfies the DU constraints, it is sufficient to consider how the  presence of $\Delta$ baryons influences the critical density of the nucleon DU reactions.
On the left panel of Fig.~\ref{fig:cut03-Udep-1} we show the critical density, $n_{\rm DU}$ and the critical NS mass for the DU reactions on nucleons, $M_{\rm DU}$, as functions of the value of the $\Delta$ potential. For potentials $U_\Delta>-95$\,MeV in KVORcut03H$\Delta\phi$ and for $U_\Delta>-93$\,MeV  in KVORcut03H$\Delta\phi\sigma$ models $n_{\rm DU}$ is not influenced by the $\Delta$s. For deeper potentials  the critical density $n_{\rm DU}$ and the corresponding star mass $M_{\rm DU}$ decrease with a decrease of the potential  and the $M_{\rm DU}$ becomes smaller than $1.5\,M_\odot$ for $U_\Delta< -109$\,MeV and $M_{\rm DU}<1.35\,M_\odot$ for $U_\Delta<-125$\,MeV.
For an unrealistically deep potential $U_\Delta < -110$\,MeV, $n_{\rm DU}$ and $M_{\rm DU}$ for KVORcut03$\Delta$, KVORcut03H$\Delta\phi$ and KVORcut03H$\Delta\phi\sigma$ models coincide with each other.

On the right panel of Fig.~\ref{fig:cut03-Udep-1} we show the maximum mass of NSs as a function of the value of the $\Delta$ potential.
For the KVORcut03$\Delta$ model $M_{\rm max}$ decreases from 2.17 $M_\odot$ at $U_\Delta =- 50$\,MeV to $2.13\,M_\odot$ at $U_\Delta\simeq -150$\,MeV but still remains well above the empirical constraint. For KVORcut03H$\Delta\phi\sigma$ and especially for KVORcut03H$\Delta\phi$ models the $U_\Delta$ dependence is  very weak.
For KVORcut03H$\Delta\phi\sigma$ model for $U_\Delta <-85$\,MeV the maximum mass slightly decreases with deepening of the potential but still remains  above the empirical constraint and for the KVORcut03H$\Delta\phi$ model the maximum star mass proves to be on the lower border of the allowed empirical constraint for all $U_\Delta$. Here, we would like to pay attention to a peculiar behaviour of $M_{\rm max}(U_\Delta)$  in the interval $-130<U_\Delta <-150$~MeV: the maximum NS mass slightly increases with the deepening of the $U_\Delta$.

\begin{figure}
\centering
\includegraphics[width=14cm]{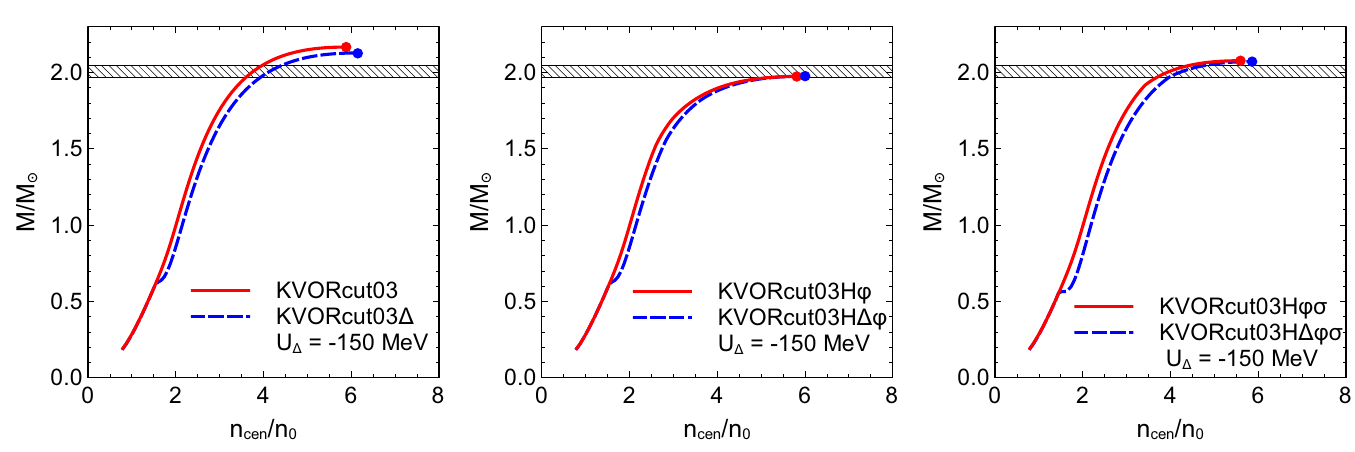}
\caption{The NS mass as a function of the central baryon density in KVORcut03,
KVORcut03$\Delta$ (left panel), KVORcut03H$\phi$, KVORcut03H$\Delta\phi$ (middle panel), KVORcut03H$\phi\sigma$, and KVORcut03H$\Delta\phi\sigma$ (right panel) models for $U_\Delta=-150$\,MeV. The $\Delta$ parameters are taken as in Eq.~(\ref{x-QCU}).
The horizontal band shows the uncertainty range for the mass of PSR J0348+0432
($2.01\pm 0.04\,M_\odot$).}
	\label{fig:cut03-Mn}
\end{figure}

\begin{figure}[t]
\centering
\includegraphics[width=14cm]{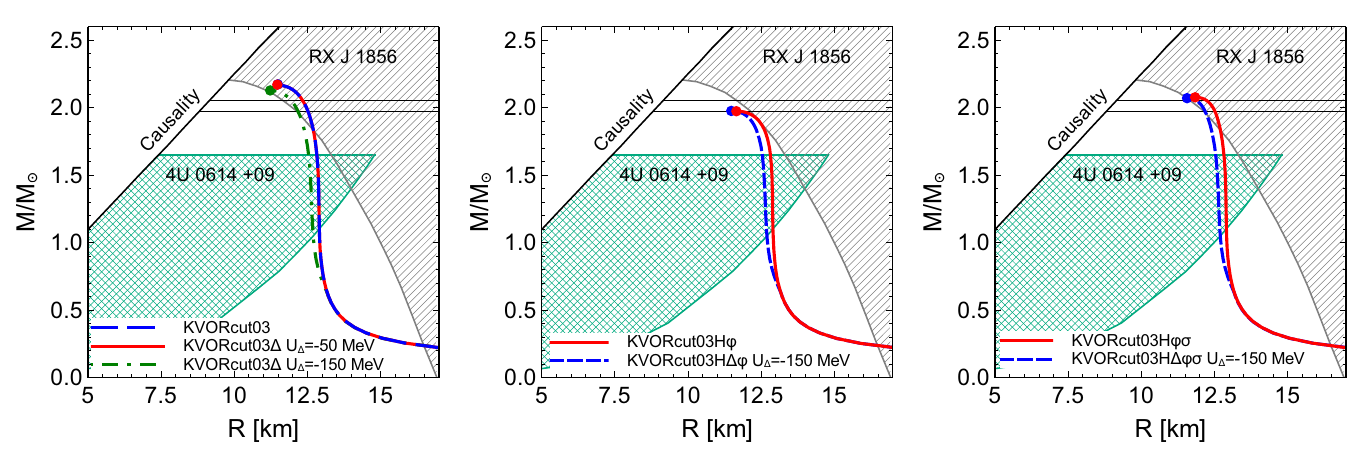}
\caption{NS mass-radius plot for the same models as in Figs.~\ref{fig:cut03-conc} and \ref{fig:cut03-Mn}  and $U_\Delta=-50$\,MeV and $-150$\,MeV together with constraints from thermal radiation of the isolated NS RX~J1856~\cite{Trumper} and from QPOs in the LMXBs 4U 0614+09~\cite{Straaten}. The $\Delta$ parameters are taken as in Eq.~(\ref{x-QCU}).  The band shows the uncertainty range  for the mass of pulsar
J0348+0432~\cite{Antoniadis:2013pzd}. For $U_\Delta=-50$\,MeV the lines for KVORcut03H$\phi$ and KVORcut03H$\Delta\phi$ models, and for KVORcut03H$\phi\sigma$ and KVORcut03H$\Delta\phi\sigma$ models  coincide since $\Delta$ do not appear.  }
\label{fig:cut03-MR}
\end{figure}

In Refs.~\cite{Drago2014,Drago:2015cea} the authors argue that the appearance of $\Delta$s in a NS with the given central density $n_{\rm cen}$ results in a notable reduction of the NS mass. We find, however, that in the KVORcut03$\Delta$ model  the star mass decreases in average by $0.002M_\odot$ at a given central density compared to that for KVORcut03 model for $U_\Delta=-50$ and by $0.02M_\odot$ for $-100$\,MeV.
To see a stronger influence of $\Delta$s on $M$ we should allow for  still stronger $\Delta$ attraction. In Fig.~\ref{fig:cut03-Mn} we show the dependence of the NS mass on the central density for $U_\Delta=-150$\,MeV for KVORcut03 and KVORcut03$\Delta$ (left panel), KVORcut03H$\phi$ and KVORcut03H$\Delta\phi$ (middle panel), and
KVORcut03H$\phi\sigma$ and KVORcut03H$\Delta\phi\sigma$ models (right panel). We see that even for unrealistically deep potential $U_\Delta=-150$\,MeV  in all cases the mass reduction does not exceed $0.1\,M_\odot$ for all values of $n_{\rm cen}$, whereas the BEM composition is more sensitive to the value of $U_\Delta$ (see Fig. \ref{fig:cut03-conc} and discussion above).

In Fig.~\ref{fig:cut03-MR} we compare the mass-radius relations for NSs calculated in the
KVORcut03 and KVORcut03$\Delta$ models   for $U_\Delta =-50$\,MeV and $-150$\,MeV (left panel), and  for $U_\Delta =-150$\,MeV in the KVOR\-cut03\-H$\phi$ and KVOR\-cut03\-H$\Delta\phi$ models (middle panel), and
in the KVOR\-cut03\-H$\phi\sigma$ and KVOR\-cut03\-H$\Delta\phi\sigma$ models (middle panel). In the latter two cases we show the results for $U_\Delta =-150$\,MeV only, since for $U_\Delta =-50$\,MeV in these models $\Delta$ baryons do not appear. We see that in the KVORcut03$\Delta$ model with $U_\Delta =-50$\,MeV the radius $R$ at fixed $M$ is practically unchanged compared to that in the KVORcut03 model. Even for $U_\Delta =-150$\,MeV in all considered models  $R$ changes only slightly (by $<0.5$\,km) for almost all  masses. The changes in $R$ at fixed $M$ are higher only for $M>2M_{\odot}$ in KVORcut03 and KVORcut03$\Delta$ models.

Concluding this section, we summarize that for the chosen realistic values of the $\Delta$ potential ($U_\Delta =-50$\,MeV) in the KVORcut03$\Delta$ model the influence of $\Delta$s is minor and in the KVORcut03H$\Delta\phi$ and KVORcut03H$\Delta\phi\sigma$ models $\Delta$s do not appear at all. The hyperons inhibit the appearance of $\Delta$s. Only for a very attractive potential $U_\Delta \sim-150$\,MeV the $\Delta$ baryons start contributing sizeably within these models.

\subsection{MKVOR*-based models}\label{sec:MKVOR}

 The equations of state obtained in the MKVOR- and MKVOR*-based models are more strongly affected by the $\Delta$ potential than the EoSs in the KVORcut-based models because the effective nucleon mass in the former two models is smaller at given density than in the latter models. Therefore, further focusing on the MKVOR*-based models we restrict by consideration of potentials $U_\Delta >-100$\,MeV.

\begin{figure}
\centering
\includegraphics[width=5cm]{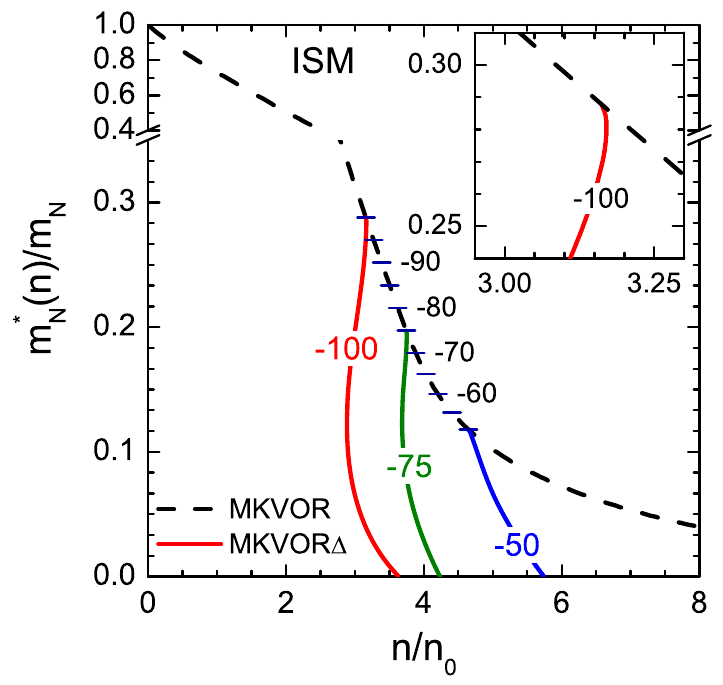}
\caption{ Effective nucleon mass as a function of the density in the ISM at various values of the $\Delta$ potential. The results obtained in the MKVOR model are shown by dashed line and the results for the MKVOR$\Delta$ model are shown by solid lines for densities where $\Delta$ baryons can exist. The values  of the potential $U_\Delta$ in MeV are indicated by labels on the lines. Horizontal ticks mark the points where solid lines branch out from dashed line for the intermediate values of $U_\Delta$.}
\label{fig:mkv-meff}
\end{figure}

In  Fig.~\ref{fig:mkv-meff}  we show the effective nucleon mass in ISM as a function of the density for MKVOR model and MKVOR$\Delta$ model for various values of $U_\Delta$. We see that the effective nucleon mass  reaches zero at some density  $n=n_{{\rm c},f=1}(U_\Delta)$.
Hence, for $n>n_{{\rm c},f=1}(U_\Delta)$ the hadron description of ISM within MKVOR$\Delta$-based models is impossible.  Thus, the density $n_{{\rm c},f=1}(U_\Delta)$ is the endpoint of our hadronic EoS for a certain $U_\Delta$.
At this point the MKVOR model should be matched with a quark model in order to proceed to higher densities. To extend the purely hadron description to higher densities
 we  minimally modify the $\omega$ sector of the MKVOR model and introduce a cut for $f>f^*_{\omega}=0.95$ according to Eq.~(\ref{etaMKVOR*}).
 The so-modified MKVOR$\Delta$ model we denote as the MKVOR*$\Delta$ model.

\begin{figure}
\centering
\includegraphics[width=14cm]{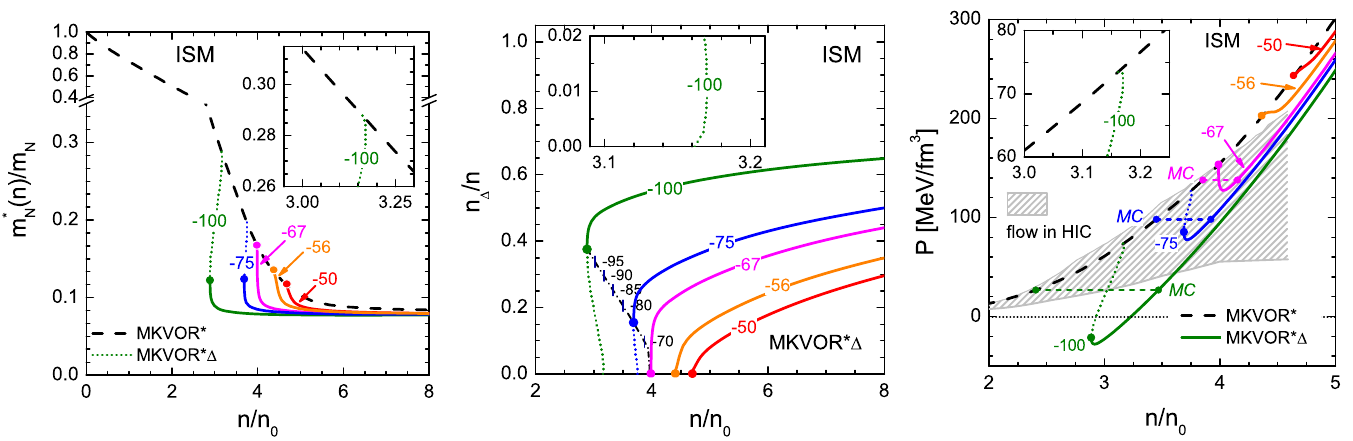}
\caption{{\em Left panel:}
Effective nucleon mass as a function of the density in the ISM for the MKVOR* model (dashed line) and the MKVOR*$\Delta$ model (dotted and solid lines) for several values of the potential $U_\Delta$ indicated by labels in MeV. Bold dots show the values of $m^*_N$ related to the critical density $n_{c,\Delta}(U_\Delta)$ at which $\Delta$ baryons may exist in the ISM. {\em Middle panel:} Concentration of $\Delta$s in the ISM as a function of the density for the  MKVOR*$\Delta$  model. Full dots show critical densities and concentrations for the appearance of $\Delta$s. The dash-dotted line connecting the full dots shows $n_{c,\Delta}(U_\Delta)$ as a function of $U_\Delta$, which variation steps are indicated by vertical bars. {\em Right panel:} Pressure as a function of the density in the ISM for MKVOR* model (dashed line) and MKVOR*$\Delta$ model (dotted and solid lines). The hatched region indicates the nucleon flow~\cite{Danielewicz:2002pu} constraints in heavy-ion collisions.}
\label{fig:mkvstar-meff}
\end{figure}

On the left panel of Fig.~\ref{fig:mkvstar-meff} we show the effective nucleon mass in ISM as a function of the density for MKVOR*  model (dashed line) and for MKVOR*$\Delta$ model (solid and dotted lines) for several values of $U_\Delta$. For $U_\Delta>-67$\,MeV, the effective mass $m_N^*$ decreases monotonously in the MKVOR*$\Delta$ model with an increase of $n$ and approaches a limiting non-zero value $m_N^*[\rm lim]\simeq 0.079\, m_N$ for large $n$. For potentials deeper than $-67$\,MeV  the curve $m_N^* (n)$ receives a back-bending segment (dotted lines) between two points with $\rmd m_N^*/\rmd n =\infty$. One of these points is explicitly marked by the bold dot in the main frame on the left panel, whereas the presence of the second point is seen only in the insertion, where the curve for $U_\Delta =-100$~MeV is shown. With a further increase of $n$, after the back-bending region, $m_N^* (n)$ decreases monotonously in MKVOR*$\Delta$ model (solid lines) tending to the same limiting non-zero value as for the MKVOR* model.

On the middle panel of Fig.~\ref{fig:mkvstar-meff} we show the $\Delta$ baryon concentrations, $n_{\Delta}$, in the MKVOR*$\Delta$ model for ISM as functions of $n$, for the same values of $U_\Delta$ as on the left panel. The back-bending region for $U_\Delta <-67$\,MeV is also manifested in this figure (dotted lines) between two points, $n_{\rm L}$ and $n_{\rm R}$ ($n_{\rm L}<n_{\rm R}$) in which $\rmd n_\Delta/\rmd n =\infty$.  One of them, $n_{\rm R}$, corresponding to a smaller $n_\Delta$ is exemplified in the figure insertion only for $U_\Delta=-100$\,MeV. The point $n_{\rm L}$ corresponds to a higher value of $n_\Delta$ and is indicated by solid dots in the main frame of Fig.~\ref{fig:mkvstar-meff}.
For densities between these points the equation $\mu_N(n,n_\Delta)=\mu_\Delta(n,n_\Delta)$, determining the $\Delta$ abundance as a function of $n$, has several solutions (two or three depending on the density).
The density $n_{\rm L}$ is the smallest density at which the $\Delta$ baryons can exist in the ISM. With the deepening of the potential $U_\Delta$ this critical density is shifted to lower values and the corresponding starting concentration of $\Delta$s increases. For densities $n>n_{\rm L}$ on the upper branch of solutions $n_\Delta(n)$, shown by solid line, $n_\Delta(n)$ increases monotonously with a density increase and the $\Delta$ concentration on this branch is the higher, the more attractive the potential $U_\Delta$ is. For $U_\Delta\ge -67$\,MeV the density points $n_{\rm R}$ and $n_{\rm L}$ coalesce and disappear, and the back-bending region disappears too.

On the right panel of Fig.~\ref{fig:mkvstar-meff} we show the pressure as a function of the density for the MKVOR* model (dashed line) and for the MKVOR*$\Delta$ model (dotted and solid lines) for densities where $\Delta$s are present for several potentials $U_\Delta$. For the MKVOR and MKVOR* models the pressure $P(n)$ starts violating the particle-flow constraint of~\cite{Danielewicz:2002pu} at $n>4.06\, n_0$ (dashed line escapes the hatched region).  We see that in the MKVOR*$\Delta$ model with $-83\,{\rm MeV} <U_\Delta <-65$\,MeV, the constraint is fulfilled for densities $n_0<n<4.5\, n_0$. This means that, if the constraint suggested in~\cite{Danielewicz:2002pu} is confirmed by subsequent more detailed analyses, this circumstance could be considered as a constraint on $U_\Delta$. For $U_\Delta >-56$\,MeV, $P(n)$ undergoes a smooth bend  in the critical point for the $\Delta$ appearance. Such a behaviour is typical for  a third-order phase transition.
Contrary, for $U_\Delta <-56$\,MeV, the curve $P(n)$ demonstrates the behaviour typical for a first-order phase transition with three solutions of the equation $P(n)=P_0={\rm const}$ in some interval of $P_0$. For $-67<U_\Delta <-56$\,MeV there exists an ordinary spinodal region with a negative incompressibility.
Interestingly, for potentials $U_\Delta$ deeper than $-67$\,MeV there appears a specific back bending of the $P(n)$ curve for densities $n_{\rm L}<n< n_{\rm R}$
with $n_{\rm L(M)}$ introduced above. Note that at these densities we have $\rmd P/\rmd n =\infty$, and $n_{\rm L}$ is marked by the dot in the main frame in the figure and the presence of the second point $n_{\rm R}$ is exemplified in the insertion. There are two narrow spinodal regions close to these points and the curve connecting these two points (dotted line) with a positive incompressibility.
A thermodynamical equilibrium between the states with and without $\Delta$s is established
along a line on the $P$--$n$ diagram connecting points of equal pressures and equal baryon chemical potentials of both states: $P(n_1^{\rm MC})=P(n^{\rm MC}_2)$ and $\mu (n_1^{\rm MC})=\mu(n_2^{\rm MC})$. These  MC lines are depicted by short dashed lines on the right panel of Fig.~\ref{fig:mkvstar-meff}. Note that back-bending behaviour of $P(n)$ has been found for ordinary RMF models in ISM at high temperatures \cite{Glendenning87}. For $U_\Delta =-91.4$\,MeV the curve $P(n)$ touches a zero line at $n=3.15\,n_0$. For $U_\Delta <-91.4$\,MeV the function $P(n)=0$ crosses zero at two values of the density for $n>n_0$. One of these zeros corresponds to an unstable state (left one), the other (right one) to a metastable state.
Note that a first-order phase transition owing to the appearance of $\Delta$s that we obtained within MKVOR*$\Delta$ model for $U_\Delta <-56$~MeV could manifest itself as an increase of the pion yield at typical energies and momenta corresponding to the $\Delta\to \pi N$ decays in heavy-ion collision experiments.

\begin{figure}[!t]
\centering
\includegraphics[width=14cm]{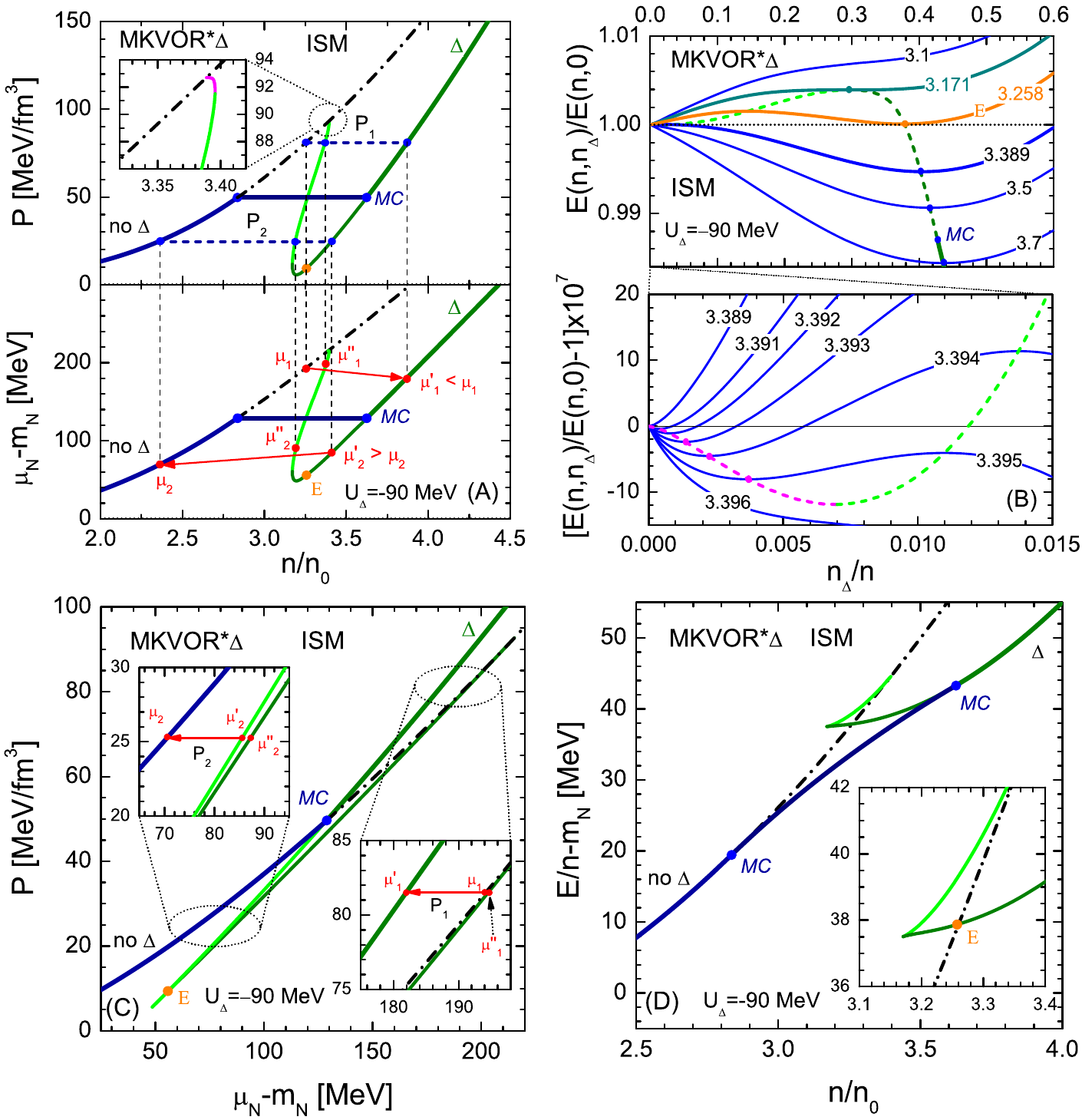}
\caption{
Paths of a first-order phase transition in the ISM for MKVOR*$\Delta$ model for $U_\Delta =-90$\,MeV illustrated in various thermodynamical quantities.
{\em Panel A:} Pressure $P(n)$ and chemical potential $\mu(n)$ as function of density
for equilibrium concentration of $\Delta$ baryons following from Eq.~(\ref{eq4muD}).
{\em Panel B:} Normalized energy density $E(n,n_\Delta)$ as a function of $\Delta$ concentration for a fixed total density indicated by labels (in $n_0$).
{\em Panel C:} Pressure as a function of the chemical potential for the equilibrium $\Delta$ concentration.
{\em Panel D:} Energy per baryon $E/n-m_N$ vs. total density for the equilibrium $\Delta$ concentration.
Line styling of the corresponding parts of the curves is the same on all panels, e.g., thick lines show the equilibrium evolution of the system through the MC.
}
\label{Fig-P-new}
\end{figure}

In the case of a usual van-der-Waals EoS there is no back bending region of the $P(n)$ curve for any
density and in the corresponding spinodal region the incompressibility is negative. In our case of the MKVOR*$\Delta$ model the usual spinodal region exists only for  potentials  $-67\,{\rm MeV}<U_\Delta <-56$\,MeV.  As we have mentioned, for $U_\Delta <-67$\,MeV besides a spinodal region there appears an unusual  back banding region, where the incompressibility is again positive (between two points in which $dP/d n =\infty$).
It is interesting to study this phenomenon in a more detail.
Therefore, in Fig.~\ref{Fig-P-new} we present various thermodynamic quantities in the phase-transition region for the MKVOR*$\Delta$ model for $U_\Delta=-90$\,MeV. For this potential the pressure is positive for any density $n>n_0$.

On panel A of Fig.~\ref{Fig-P-new}
we show $P(n)$ and $\mu_N (n)$. On panel B we illustrate the dependence of the energy density on the $\Delta$ concentration. On panel C we present the $P(\mu)$ dependence. On panel D the energy per particle is plotted as a function of the density. All these quantities are calculated for the equilibrium concentration of $\Delta$ baryons. Bold curves on all panels demonstrate the path of the system being at equilibrium. The horizontal segments on panels A and C
corresponding to $P=P^{\rm MC}=49.6\,{\rm MeV/fm^3}$ and $\mu_N =\mu_{N}^{\rm MC}=1070$\,MeV are the MC lines, on panel C they correspond to a point labeled MC.
The difference in the energy per particle and the $\Delta$ concentration between the end points on the MC line can be inferred from position of the MC points on panels D and B, respectively.
Labels ``$\Delta$" and ``no $\Delta$" mark the parts of the equilibrium curve (thick solid lines) with and without $\Delta$ baryons, respectively.  Along the MC line on panel A one can speak only about an averaged density of the matter, which varies between $n_1^{\rm MC}=2.84\,n_0$ and $n_2^{\rm MC}=3.63\,n_0$ according to the equation $n=\bar{n}=n_1^{\rm MC}(1-f_\Delta)+f_\Delta n_2^{\rm MC}$, where $f_\Delta$ is the relative fraction of the volume occupied by the ``$\Delta$'' phase. The $\Delta$ concentration rises from $x_{\Delta,1}=0$ in the beginning of the MC line to $x_{\Delta,2}=0.43$
according to the equation $\bar{x}_\Delta=x_{\Delta,2} (n_2^{\rm MC}/\bar{n})(\bar{n}-n_1^{\rm MC})/(n_2^{\rm MC}-n_1^{\rm MC})$.
To clarify the balance between the phases with and without $\Delta$s beyond the MC line, let us consider the system at two fixed pressures $P=P_1>P^{\rm MC}$ and  $P=P_2<P^{\rm MC}$ (short-dash lines on panel A).  In the former case the system, being initially placed
in state 1 without $\Delta$ (on dash-dotted line) or state $1''$ with a low $\Delta$ concentration should after a while come to stable state $1'$ (on thick solid line) with an equilibrium concentration of $\Delta$, since $\mu''_1>\mu_1>\mu'_1$. The corresponding chemical potentials are indicated also on graphs $\mu(n)$ and $P(\mu)$. The state 1 with $P_1$ and  $\mu_1$  corresponds to the state usually named as an ``overheated gas''.
Similarly, if at the fixed pressure $P_2$ one starts in state $2'$  ($P_2,\,\mu'_2$) on the ``$\Delta$'' part of thick solid curve with a large $\Delta$ concentration, the system will evolve to state 2 without $\Delta$s, since $\mu'_2>\mu_2$. The same happens if one starts in an intermediate state $2''$ on the back-bent piece of solid line since $\mu''_2>\mu'_2>\mu_2$. Continuing an analogy with the ordinary liquid-gase phase transition, state $2'$ can be named as a ``supercooled liquid". In equilibrium $P(\mu)$ should be maximum, hence the system undergoing a first-order phase transition follows in equilibrium the path shown by thick lines on panel C.

To illustrate how the system chooses the appropriate concentration of $\Delta$ baryons we consider energy density of the system $E(n,n_\Delta)$ as a function of $n_\Delta$ for various fixed values of the total density $n$. On panel B we plot the dimensionless ratio $E(n,n_\Delta)/E(n,0)$ to get rid off the common $n$ dependence. For densities $n\lsim 3.171\,n_0$ the curve is monotonously increasing with an increase of $n_\Delta$ with the global minimum at $n_\Delta =0$ that corresponds to the ``no $\Delta$'' curve on panel A.  The density $n\approx 3.171\,n_0$ corresponds to the point $\rmd P/\rmd n=\infty$ and $\rmd \mu/\rmd n=\infty$ on panel A.  For $3.171\,n_0\lsim n \lsim 3.258\,n_0$, the curve $E(n={\rm const}, n_\Delta)$ has two local extrema in which $\partial E(n,n_\Delta)/\partial n_\Delta=\mu_\Delta-\mu_N=0$ and, therefore, they correspond to the chemical equilibrium between $\Delta$ and nucleons in ISM [see Eq.~(\ref{eq4muD})]. One extremum (for a smaller value of $n_\Delta$) is the local maximum of the energy density and the second one is the local minimum. The energy density at this minimum is, however, still higher then for $n_\Delta=0$, so the state
without $\Delta$s is energetically preferable, see also panel D where the ``nose'' formed by two solutions with $n_\Delta\neq 0$ is above dash-dotted line for $n_\Delta=0$ at $n<3.258\,n_0$.
At $n\approx 3.258\,n_0$ the energy densities of the ISM without $\Delta$s and with the $\Delta$ concentration $n_\Delta/n\approx0.38$ become equal. This situation is shown on panel B by the curve labeled with E and by the dots with label E on panels A, C, and D.
For all higher densities the ``$\Delta$" state is preferable since its energy is smaller, and $\Delta$ concentration increases with a growth of the density. On panel B in the density interval $3.389\,n_0<n<3.3957\,n_0$ there exist two local minima of $E$, one at a tiny concentration $n_\Delta/n\lsim 0.005$ (see lower graph on panel D) and the other much deeper one at $n_\Delta/n\sim 0.4$.
The former state is metastable and the latter is stable. This density range corresponds to the spinodal instability region shown in the insertion on the $P(n)$ graph on panel A.
Dashed line connecting extrema of $E(n={\rm const},n_\Delta)$ on panel B is related to the back-bending pieces on panel A.
For densities $n\gsim 3.3957\,n_0$ there remains only one global minimum at large $\Delta$ concentrations. On panel D the curve  between two MC points is determined by the condition $\bar{\cal E}_\Delta={\cal E}_1+ {\cal E}_2 (n_2^{\rm MC}/\bar{n})(\bar{n}-n_1^{\rm MC})/(n_2^{\rm MC}-n_1^{\rm MC})$, where ${\cal E}_1=(E(n)/n)|_{n_1^{\rm MC}}$, on the curve ``no $\Delta$", ${\cal E}_2=(E(n))|_{n_2^{\rm MC}}$, on the curve ``$\Delta$".


\begin{figure}
\centering
\includegraphics[width=14cm]{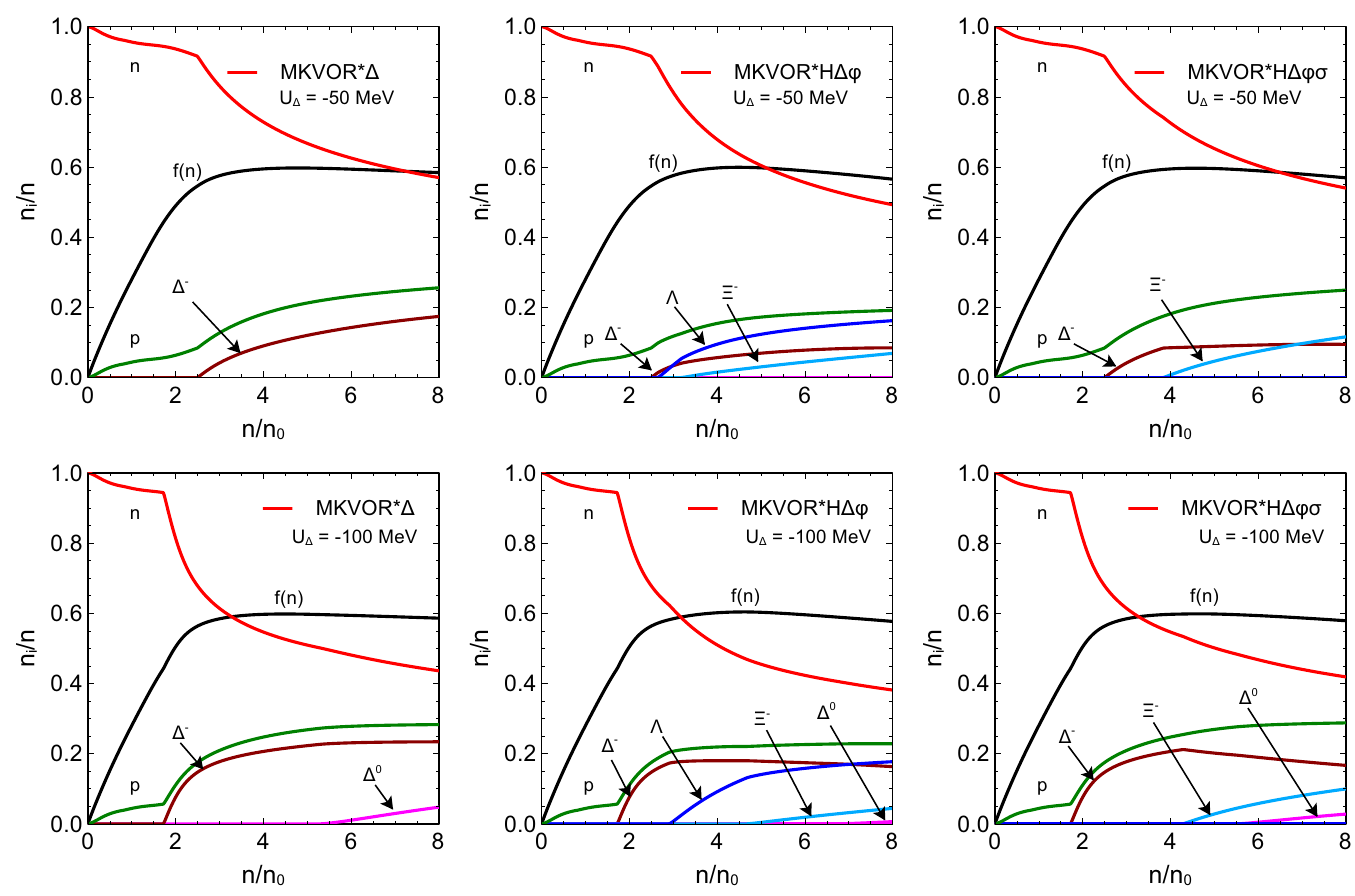}
\caption{Baryon concentrations and magnitude of the scalar field, $f(n)$, in the BEM for the
MKVOR*$\Delta$, MKVOR*H$\Delta\phi$, and MKVOR*H$\Delta\phi\sigma$ models
for  $U_{\Delta}=-50$ MeV (upper row) and $U_{\Delta}=-100$~MeV (lower row). The results are obtained with $\Delta$ parameters taken as in Eq.~(\ref{x-QCU}). }
\label{fig:mkv-conc}
\end{figure}

In BEM all  results for the MKVOR- and MKVOR*-based models coincide.
In Fig.~\ref{fig:mkv-conc} we demonstrate $f(n)$ and baryon concentrations in the MKVOR*$\Delta$, MKVOR*H$\Delta\phi$ and MKVOR*H$\Delta\phi\sigma$ models in BEM for two values of the $\Delta$-potential: $U_\Delta = - 50$ MeV and $U_\Delta = -100$ MeV. In all  these models $f(n)$ first increases with an increase of the density and for $n\gsim 3n_0$ becomes approximately constant (about 0.6).
In the MKVOR*$\Delta$ model $\Delta^-$s appear at density $2.51\, n_0$  for $U_\Delta = - 50$ MeV and at $1.74 \, n_0$ for $U_\Delta = -100$ MeV. Then the $\Delta^-$ concentration increases significantly with an increase of $n$. In both MKVOR*H$\Delta\phi$ and MKVOR*H$\Delta\phi\sigma$ models $\Delta^-$s appear at smaller densities than hyperons but their presence does not change substantially the NS composition compared to the case without $\Delta$s, cf. Fig.~25 in~\cite{Maslov:2015wba}.
With an increase of the $\Delta$ attraction from $-50$\,MeV to $-100$\,MeV we observe in all models  a decrease of the critical density $n_{\rm c,\Delta^-}$ from $\sim 2.5\,n_0$ to $\sim 1.7\,n_0$. In the MKVOR*H$\Delta\phi$  model with a density increase there appear first $\Lambda$ and then $\Xi^-$ hyperons. The critical densities of their appearance increase with a decrease of the $U_\Delta$.
In the MKVOR*H$\Delta\phi\sigma$  model  only $\Xi^-$ hyperons  arise.
For $U_\Delta = - 100$ MeV in all models there appears also a small fraction of $\Delta^0$s
in the centers of the most massive NSs.

\begin{figure}[!t]
\centering
\includegraphics[width=10cm]{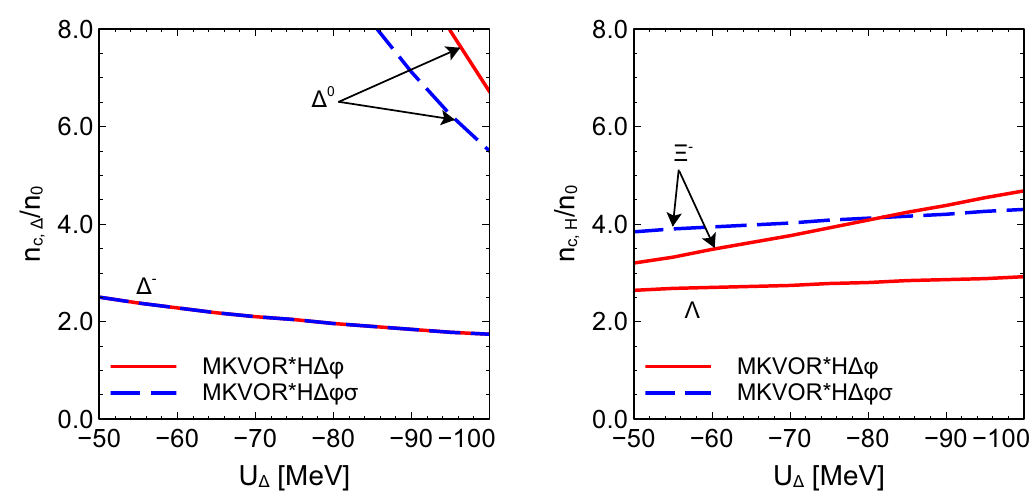}
\caption{The critical density for the appearance of $\Delta$ baryons (left panel) and  hyperons (right panel) in BEM as a function of the $\Delta$ potential for the MKVOR*H$\Delta\phi$, and MKVOR*H$\Delta\phi\sigma$ models with the $\Delta$ parameters given by Eq.~(\ref{x-QCU}).} \label{fig:mkv-Udep-nc}
\end{figure}

In Fig.~\ref{fig:mkv-Udep-nc} we demonstrate the dependence of the critical densities for the appearance of $\Delta$ baryons (left panel) and hyperons (right panel) on the value of the $\Delta$ potential. In the MKVOR*$\Delta$-based models the critical density for $\Delta^-$ baryons depends much weaker on $U_\Delta$ than that in the KVORcut03$\Delta$-based  models and is systematically smaller, cf. Fig.~\ref{fig:cut03-Udep}. The critical densities for $\Delta^{0}$ are also smaller in the MKVOR*$\Delta$-based models. $\Delta^+$ and  $\Delta^{++}$ baryons do not appear in all models even in most massive NSs but could arise, if $U_{\Delta}$ were deeper. The early appearance of $\Delta^-$s in MKVOR*$\Delta$-based models shifts $n_{\rm c,\Lambda}$ and $n_{\rm c,\Xi^-}$ to higher values the stronger, the deeper the $U_\Delta$ potential is.

\begin{figure}[!b]
\centering
\includegraphics[width=10cm]{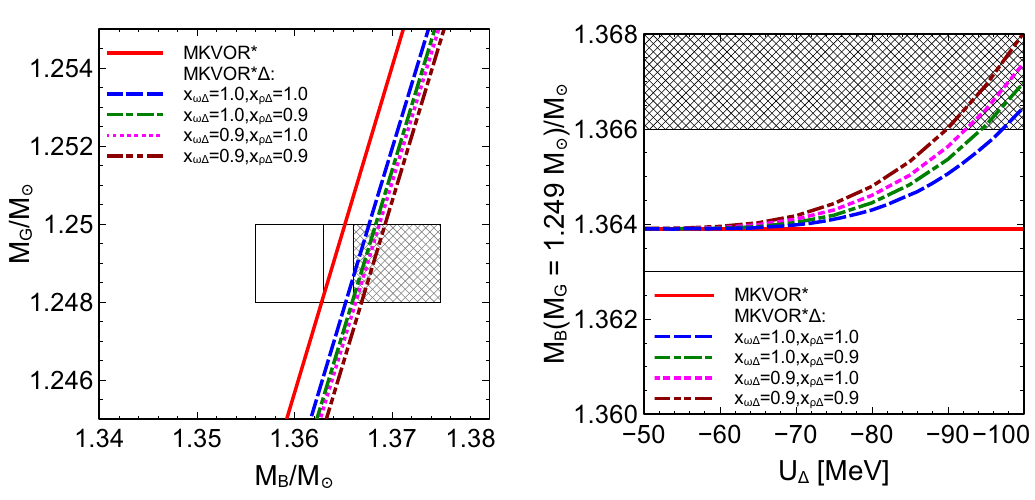}
\caption{{\em Left panel:} Gravitational-baryon NS mass constraint for MKVOR* and
MKVOR*$\Delta$ models.   The double-hatched rectangle is the constraint for the pulsar J0737-3039(B) \cite{Podsiadlowski}. The two empty rectangles show the variation of the constraint, when the assumed loss of the baryon mass during the progenitor-star collapse amounts to $0.3\% M_\odot$ and $1\% M_\odot$. {\em Right panel:} Baryon mass as a function of the $\Delta$
potential for the NS  with $M_{\rm G} = 1.249M_{\odot}$ for the MKVOR*$\Delta$ model. Double-hatched and empty bands show the corresponding experimental constraints. } \label{fig:pods}
\end{figure}

Studies of pulsar B in the double pulsar system J0737-3039~\cite{Podsiadlowski} suggested a test of the nuclear matter EoS provided a formation mechanism of the PSRJ0737-3039 system and the assumption of a negligible baryon loss of companion B during its creation are valid.
In Fig.~\ref{fig:pods} we show the gravitational mass $M_{\rm G}$ versus the baryon mass $M_{\rm B}$ of a NS. The double-hatched rectangle (left panel) and band (right panel) show the constraint from~\cite{Podsiadlowski}.  The two empty rectangles on the left panel show the allowed variation of the constraint due to the assumed loss of the baryon number during the progenitor star collapse equal to $0.3\% M_{\odot}$ (see the corresponding empty band on the right panel) and to $1\% M_{\odot}$. Approximately the same constraint box (from $0.3\% M_{\odot}$ to $1\% M_{\odot}$) was proposed in the work~\cite{Kitaura:2005bt}, which found in their model that the mass loss of the collapsing O–Ne–Mg core during the explosion leaves the NS with a baryon mass of $M=1.36 \pm 0.002M_{\odot}$. However, many EoSs do not satisfy even this weaker constraint, see Ref.~\cite{Klahn:2006ir}. The KVORcut03 curve matches marginally this ``weak" constraint, cf. Fig.~17 in~\cite{Maslov:2015wba}. Note that curves for all KVORcut03-based models (with  inclusion of hyperons and $\Delta$s) for $U_\Delta >-100$ MeV coincide with the curve for KVORcut03 model.
The MKVOR model fits marginally the ``strong" constraint (the curve touches the left boundary of the hatched box, cf. ~\cite{Maslov:2015wba}).
For the MKVOR*$\Delta$ model the agreement with the strong constraint is improved, and the better, the more attractive the assumed $\Delta$ potential is. Similar behaviour was observed also in Ref.~\cite{Drago:2015cea}. We also allowed for a variation $x_{\omega\Delta}$ and $x_{\rho\Delta}$ in limits $0.9\leq x_{\omega\Delta},x_{\rho\Delta}\leq 1$. This dependence is shown in the figure.

\begin{figure}[!b]
\centering
\includegraphics[width=14cm]{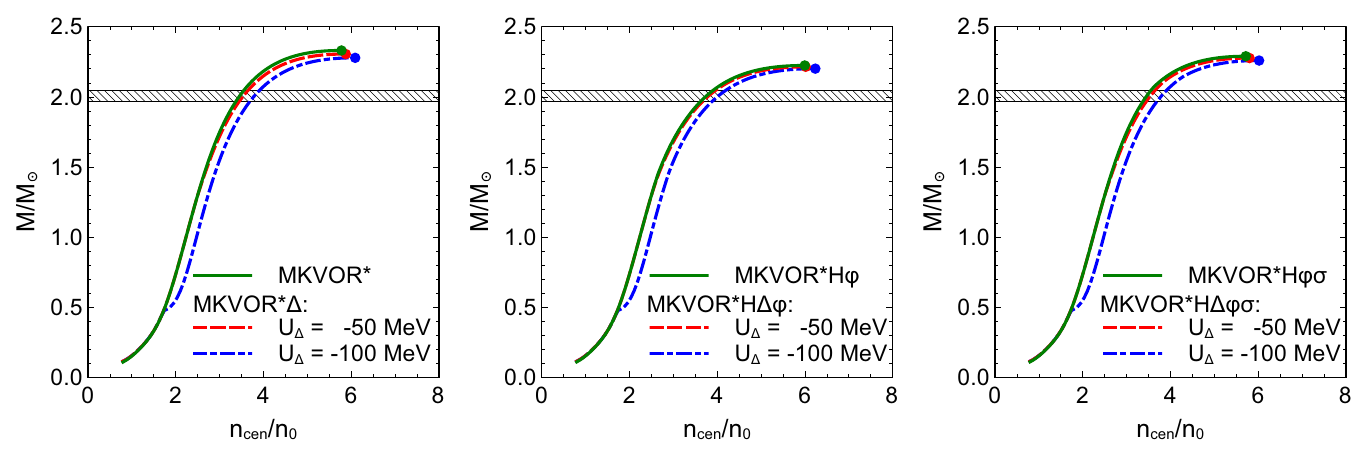}
\caption{NS mass as a function of the central baryon density in the MKVOR*,
MKVOR*$\Delta$ (left panel), MKVOR*H$\phi$ and MKVOR*H$\Delta\phi$ (middle panel), and MKVOR*H$\phi\sigma$ and MKVOR*H$\Delta\phi\sigma$ (right panel) models with
the $\Delta$ parameters taken as in (\ref{x-QCU}) for $U_{\Delta}=-50$\,MeV (solid lines) and $-100$\,MeV (dashed lines).
The horizontal band on the right panel shows the uncertainty range for the mass of PSR J0348+0432
($2.01\pm 0.04\,M_\odot$).
}
\label{fig:mkv-Mn}
\end{figure}

Figure~\ref{fig:mkv-Mn} shows the NS masses as a functions of the central density with and without $\Delta$s for MKVOR*$\Delta$ , MKVOR*H$\Delta\phi$, and MKVOR*H$\Delta\phi\sigma$ models. Despite the presence of $\Delta$s affects the NS composition substantially, the star mass
changes rather weakly. For a realistic value of the $\Delta$ potential, $U_\Delta=-50$\,MeV, the decrease of the NS mass is tiny. For a deep $\Delta$ potential, $U_\Delta=-100$\,MeV, a change of  the NS mass does not exceed $0.2\,M_\odot$. The maximum NS mass  changes even less, by $\lsim 0.05$\,$M_\odot$ only, so the maximum mass constraint is safely fulfilled even after the inclusion of $\Delta$ baryons and hyperons.

\begin{figure}\centering
\includegraphics[width=10cm]{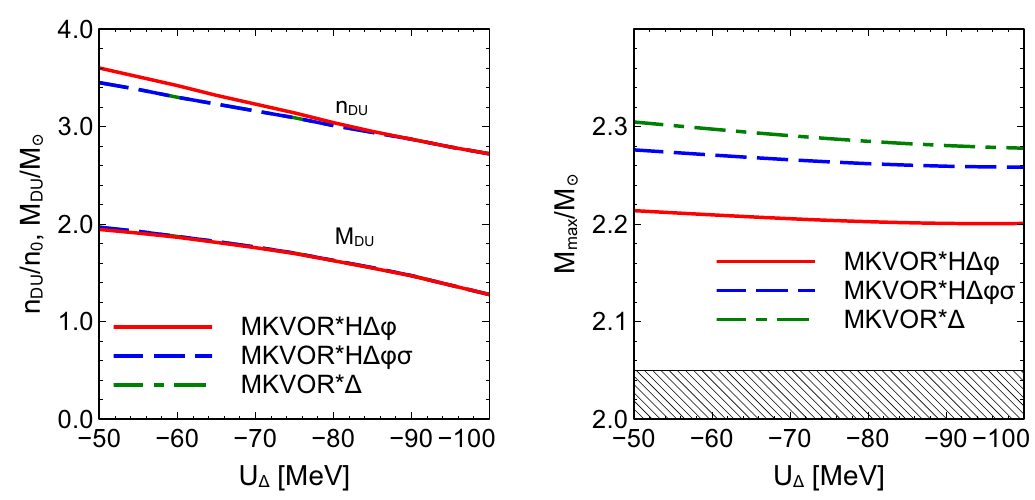}
\caption{The critical density and the critical NS mass for the DU reactions on nucleons (left panel) and the maximum masses of the NSs (right panel) as functions of the $\Delta$ potential for the MKVOR*$\Delta$, MKVOR*H$\Delta\phi$, and MKVOR*H$\Delta\phi\sigma$ models with  $\Delta$ parameters given by Eq.~(\ref{x-QCU}).
Lines for  MKVOR*$\Delta$ and MKVOR*H$\Delta\phi\sigma$ models coincide.
The horizontal band on the right panel shows the uncertainty range for the mass of PSR J0348+0432
($2.01\pm 0.04\,M_\odot$).
} \label{fig:mkv-Udep-nd-mm}
\end{figure}

The critical density and the critical NS mass of the DU reactions on nucleons in BEM are shown on the left panel of Fig.~\ref{fig:mkv-Udep-nd-mm} as functions of the $\Delta$ potential. The general trend is the same as for the KVORcut03-based models: the deepening of the $U_\Delta$ potential leads to a larger proton concentration and an earlier start of the DU reactions on nucleons. The DU constraint $M_{\rm DU}>1.35\,M_\odot$ proves to be fulfilled for $U_\Delta \gsim -96$\,MeV, the constraint $M_{\rm DU}>1.5\,M_\odot$ holds for $U_\Delta \gsim -88$\,MeV.
 As seen on the right panel of Fig.~\ref{fig:mkv-Udep-nd-mm},
the maximum mass of the NS decreases only slightly with a deepening of the potential $U_\Delta$ and remains substantially larger than the maximum among well-measured masses of the pulsars
($2.01\pm 0.04\,M_\odot$ for PSR J0348+0432).

\begin{figure}[!b]
\centering
\includegraphics[width=14cm]{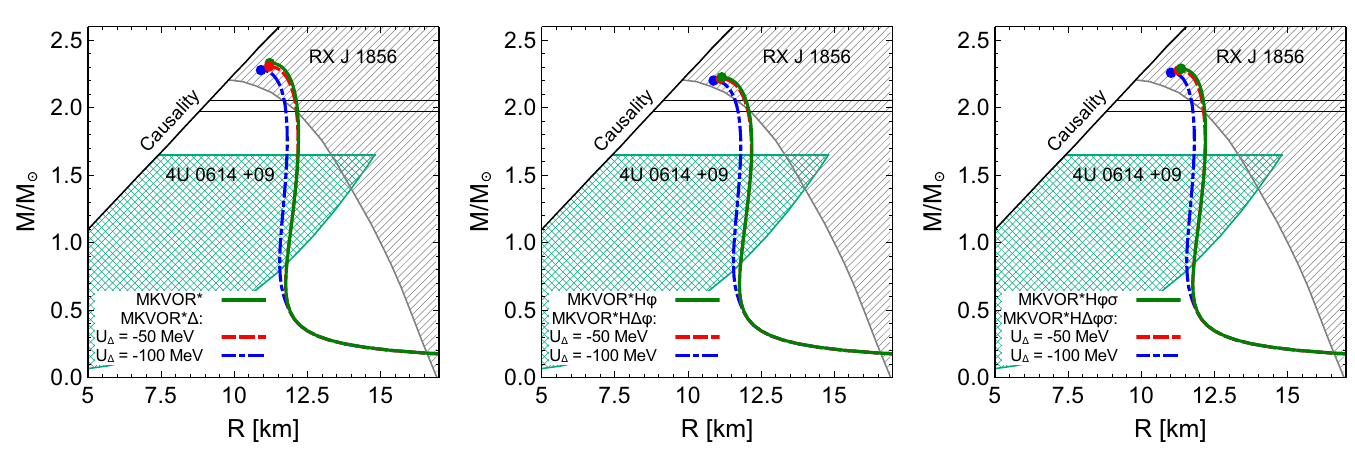}
\caption{NS mass as a function of radius for
MKVOR*, MKVOR*$\Delta$ (left panel), MKVOR*H$\phi$, MKVOR*H$\Delta\phi$ (middle panel), MKVOR*H$\phi\sigma$ and MKVOR*H$\Delta\phi\sigma$ (right panel) models with
the $\Delta$ parameters taken as in Eq.~(\ref{x-QCU}), for $U_{\Delta}=-50$\,MeV and $-100$\,MeV.
The empirical constraints are the same as in Fig.~\ref{fig:cut03-MR}. }
\label{fig:mkv-Udep-mr}
\end{figure}

Finally, Fig.~\ref{fig:mkv-Udep-mr} shows that the inclusion of $\Delta$s in MKVOR*-based models with or without hyperons does not change noticeably the mass-radius relation for NSs for $U_\Delta=-50$\,MeV. For $U_\Delta=-100$\,MeV the radius of the NS with the mass $1.5\,M_\odot$ decreases by $\sim $ 0.5\,km.

Concluding, the MKVOR*-based models with $\Delta$ baryons included with a realistic value for the $\Delta$ potential, $U_\Delta=-50$\,MeV, remain conform to astrophysical constraints as  the models without $\Delta$s.
In ISM the influence of $\Delta$s on the EoS proves to be stronger than in BEM, since in the ISM the effective baryon mass is smaller then in the BEM at the same baryon density.

\section{Additional variation of $\Delta$ parameters}\label{sec:variation}

The relation $g_{\omega\Delta}=g_{\rho\Delta}$ that follows from SU(6) symmetry  can be relaxed if one assumes  SU(3) symmetry. The SU(3) symmetrical Lagrangian involving the baryon decuplet $\Delta^{abc}_\nu$ and vector-meson nonet $(V_\mu)^a_b$ ($a,b,c=1,2,3$ are the indices in the SU(3) flavor space) has only two terms with a vector coupling
$
\mathcal{L}_{\Delta V}=
g_0\left(\bar{\Delta}_{acd}^\nu\gamma^\mu \Delta^{acd}_\nu\right) (V_\mu)^b_b+g_1\left(\bar{\Delta}_{acd}^\nu\gamma^\mu \Delta^{bcd}_\nu (V_\mu)^a_b  \right)
$, where the summation over the indices is implied.  With the standard definitions of SU(3) multiplets as, e.g. in~\cite{LutzK2002}, we find the relations
\begin{eqnarray}
g_{\om\Delta}=g_1+2g_0\,,\quad g_{\rho\Delta}=\frac23 g_1\,,\quad g_{\phi\Delta}=\sqrt{2}\, g_0\,.
\end{eqnarray}
Taking into account the Iizuka-Zweig-Okubo suppression~\cite{Okubo} of the $\phi$ meson coupling to not strange baryons and requiring, therefore, $g_{\phi\Delta}=0$ we find the relation $g_{\rho\Delta}=\frac23 g_{\om\Delta}$. This relation can be rewritten as
\begin{eqnarray}
 x_{\rho\Delta}=\frac23 x_{\om\Delta}\frac{C_{\om}m_\om}{C_{\rho}m_\rho}
\,.
\label{x-QCUR}
\end{eqnarray}
 With the parameters for the models from Tables~\ref{tab:param-KVORcut03} and \ref{tab:param-MKVOR} we get $x_{\rho \Delta}=0.63\,x_{\om \Delta}$ for the KVORcut03 model and $x_{\rho \Delta}=0.87\,x_{\om \Delta}$ for the MKVOR model instead of the relation $x_{\om\Delta}=x_{\rho \Delta}=1$ that we used exploiting the SU(6) symmetry.
Therefore, to check a sensitivity of the results to these poorly known parameters we allow now for a variation of $x_{\om\Delta}$, $x_{\rho \Delta}$ near unity.

\begin{figure}
\centering
\includegraphics[width=10cm]{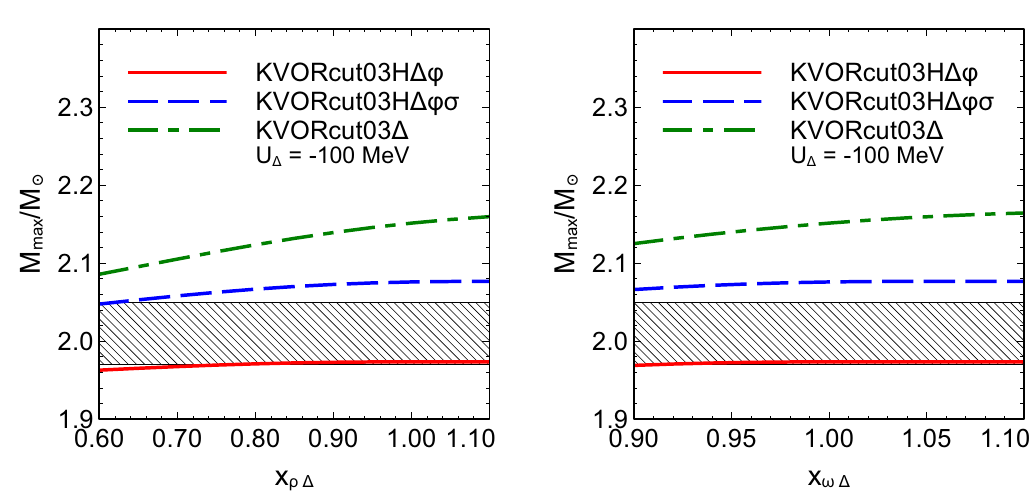}
\caption{
Maximum NS mass as a function of the parameter $x_{\rho\Delta}$ at $x_{\om\Delta}=1$ and $U_{\Delta}=-100$\,MeV (left panel) and of the parameter $x_{\om\Delta}$ at $x_{\rho\Delta}=1$ and $U_{\Delta}=-100$\,MeV (right panel) for KVORcut03-based models.
The horizontal band  shows the uncertainty range for the mass of PSR J0348+0432
($2.01\pm 0.04\,M_\odot$).
}
\label{fig:cut03xr-xom}
\end{figure}

In Fig. \ref{fig:cut03xr-xom} we show   the maximum NS mass as a function of the parameter $x_{\rho\Delta}$ at $x_{\om\Delta}=1$ and $U_{\Delta}=-100$\,MeV (left panel) and of the parameter $x_{\om\Delta}$ at $x_{\rho\Delta}=1$ and $U_{\Delta}=-100$\,MeV (right panel) for KVORcut03$\Delta$-based models. We see that for models with hyperons --- KVORcut03H$\Delta\phi$ and KVORcut03H$\Delta\phi\sigma$ --- the maximum NS mass is rather insensitive to the variation of $x_{\om\Delta}$ and $x_{\rho\Delta}$, whereas the maximum NS mass  in the KVORcut03$\Delta$ model is more sensitive to these variations. We proved that for $U_{\Delta}=-50$\,MeV, $\Delta$ baryons do not appear in the KVORcut03H$\Delta\phi$ and KVORcut03H$\Delta\phi\sigma$ models, and the dependence on $x_{\om\Delta}$ and $x_{\rho\Delta}$ parameters in the KVORcut03$\Delta$ model is weaker for $U_{\Delta}=-50$\,MeV than for $U_{\Delta}=-100$\,MeV. For all relevant values of the coupling parameters the KVORcut03$\Delta$ and KVORcut03H$\Delta\phi\sigma$ models do appropriately fulfill the maximum NS mass constraint. The  KVOR\-cut03\-H$\Delta\phi$ and KVORcut03H$\phi$ models without $\Delta$s fulfill this constraint only marginally.

\begin{figure}
\centering
\includegraphics[width=10cm]{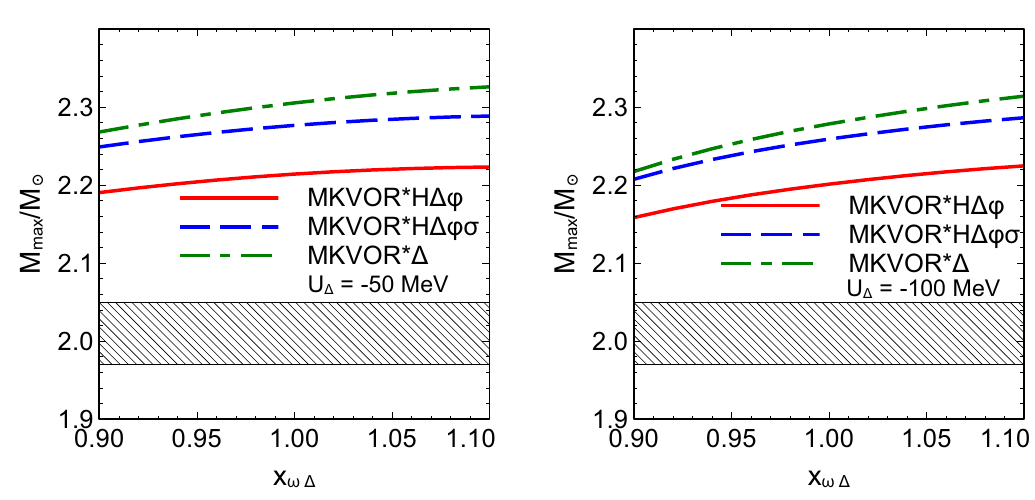}
\caption{Maximum NS mass as a function of the parameter $x_{\om\Delta}$ at $x_{\rho\Delta}=1$ for $U_{\Delta}=-50$\,MeV (left panel) and for $U_{\Delta}=-100$\,MeV (right panel) for MKVOR*-based models.
The horizontal band  shows the uncertainty range for the mass of PSR J0348+0432
($2.01\pm 0.04\,M_\odot$).
}
\label{fig:mkv-xo}
\end{figure}

\begin{figure}
\centering
\includegraphics[width=10cm]{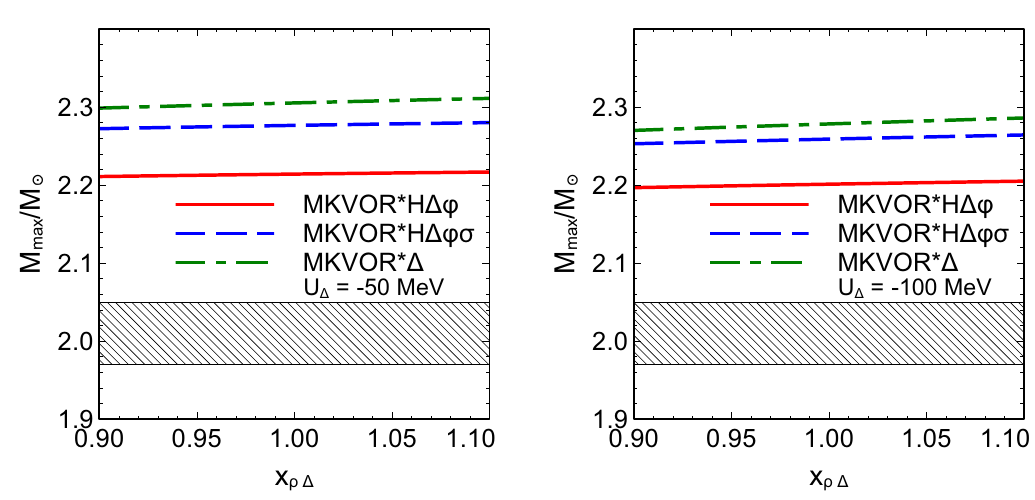}
\caption{Maximum NS mass as a function of the parameter $x_{\rho\Delta}$ at $x_{\om\Delta}=1$ for $U_{\Delta}=-50$\,MeV (left panel) and for $U_{\Delta}=-100$\,MeV (right panel) for MKVOR*-based models.
The horizontal band  shows the uncertainty range for the mass of PSR J0348+0432
($2.01\pm 0.04\,M_\odot$).
}
\label{fig:mkv-xr}
\end{figure}

In Fig. \ref{fig:mkv-xo}   we show the maximum NS mass as a function of the parameter $x_{\om\Delta}$ at $x_{\rho\Delta}=1$ for $U_{\Delta}=-50$\,MeV (left panel) and for $U_{\Delta}=-100$\,MeV (right panel) for the MKVOR-based models. In Fig. \ref{fig:mkv-xr}   we demonstrate the maximum NS mass as a function of the parameter $x_{\rho\Delta}$ at $x_{\om\Delta}=1$ for $U_{\Delta}=-50$\,MeV (left panel) and for $U_{\Delta}=-100$\,MeV (right panel) for MKVOR*-based models. Here, all the models MKVOR*$\Delta$, MKVOR*H$\Delta\phi$, MKVOR*H$\Delta\phi\sigma$ appropriately fulfill the maximum NS mass constraint in the whole range of varied parameters.


\section{Conclusion}

In~\cite{Maslov:2015msa,Maslov:2015wba} we proposed  several relativistic mean-field (RMF) models with scaled hadron masses and coupling constants depending self-consistently on the scalar mean-field.  These models are the extensions of the KVOR model proposed in~\cite{Kolomeitsev:2004ff} and then successfully tested in~\cite{Klahn:2006ir} against various experimental constraints. Within these models all hadron masses are assumed to decrease universally with the scalar field growth, whereas the meson-nucleon coupling constants can vary differently. The aim in~\cite{Maslov:2015msa,Maslov:2015wba} was to construct an RMF model, which satisfies presently  known experimental constraints put on the equation of state (EoS) from various analyses of atomic nuclei, heavy-ion collisions and pulsars. Especial  challenge is that
the EoS of the beta-equilibrium  matter (BEM) should be sufficiently stiff to support the existence of neutron stars (NSs) with the mass $>2\,M_\odot$ and, simultaneously, the EoS of the isospin symmetrical matter (ISM) should respect the constraint derived from flows of particles produced in heavy-ion collisions~\cite{Danielewicz:2002pu}. We have exploited a novel mechanism of stiffening of the EoS in the framework of a RMF model described in~\cite{Maslov:cut}  (named the cut-mechanism), which assumes a limitation of the growth of the scalar field at densities above some chosen one. It is achieved by a special choice of the scaling functions.

In the given work we focused on extensions of the models KVORcut03 and MKVOR, which we have formulated in \cite{Maslov:2015wba}.
The  KVORcut03 model exploits the cut mechanism in the $\om$ sector, whereas MKVOR model uses the cut mechanism in the $\rho$ sector. In previous works ~\cite{Maslov:2015msa,Maslov:2015wba} we allowed for occupation of the hyperon Fermi seas in dense BEM.  We exploited the  choice of the couplings of the hyperons (H) with  $\omega$, $\rho$ and $\phi$ fields in vacuum  according to SU(6) symmetry. The $H\sigma$ coupling was constrained  by the experimental information on the hyperon potentials in nuclei.
We demonstrated in \cite{Maslov:2015wba} that with two choices for inclusion of hyperons (in the KVORcut03H$\phi$ and KVORcut03H$\phi\sigma$, and MKVORH$\phi$ and MKVORH$\phi\sigma$ models)
the experimental constraints on the EoS continue to be fulfilled. By this we resolved the so-called ``hyperon puzzle" in the framework of thus constructed RMF models: the EoS satisfies the experimental constraint on the minimal value of the maximum mass of the NSs. However, we disregarded in mentioned works a possibility of the filling of Fermi seas of $\Delta$ isobars. As argued in~\cite{Drago2014,Cai:2015hya,Drago:2015cea} besides the hyperon puzzle there exists the similar $\Delta$ puzzle. Therefore, in the present paper we incorporate $\Delta$s in our models.

The coupling constants of the $\Delta$ resonances are poorly constrained empirically,
due to unstable nature of the $\Delta$ particles and the complicated pion-nucleon
dynamics in medium. Basing on SU(6) symmetry relations, we exploited the universal choice of the couplings of $\Delta$s with $\omega$ and $\rho$ fields in vacuum. The $\sigma\Delta$ coupling was constrained by choosing  a value for the $\Delta$ potential at nuclear saturation density $U_\Delta (n_0)$ where  $n_0\simeq 0.16$\,fm$^{-3}$. We varied the value $U_\Delta (n_0)$ in broad limits.  Then we also allowed for a variation of the $\Delta$ coupling constants with  $\omega$ and $\rho$ fields. The $\phi\Delta$ coupling  is held zero.

We demonstrated that within the KVORcut03$\Delta$ model $\Delta$s do not appear in the ISM up to extremely high densities if we choose an appropriate value of the $\Delta$ potential, $U_\Delta (n_0)=U_N (n_0) \sim -50$~MeV, cf. Fig. \ref{fig:ncD-ISM-cut03}.
The critical density for the appearance of $\Delta$s decreases, if we allow for a more attractive potential $U_\Delta (n_0)$ (that is not excluded by the data) but even for the unrealistically large attraction with $U_\Delta =-150$~MeV, the critical density of the appearance of $\Delta$s, $n_{c,\Delta}$, remains as high as 5$n_0$. In the BEM for the chosen realistic value of the potential, $U_\Delta (n_0)=-50$~MeV, $\Delta^-$ baryons arise only at densities $n> 5\, n_0$, cf. Fig. \ref{fig:cut03-conc}. Other $\Delta$ species ($\Delta^0$ and $\Delta^+$) do not appear  up to maximum densities reachable in NS interiors.

In the presence of hyperons $\Delta$ baryons do not appear in the KVOR\-cut03\-H$\Delta\phi$ and
KVORcut03H$\Delta\phi\sigma$ models for $U_\Delta (n_0)=-50$~MeV but could arise if $U_\Delta (n_0)$ were more attractive. Therefore, we artificially increased the $\Delta$-nucleon attraction allowing $U_\Delta (n_0)$ to vary within the range of $-$(50\mbox{--}150)\,MeV to investigate how it could affect the EoS in all our KVORcut03-based models.  The  critical value of the NS mass for the begining of the DU reactions on nucleons proves to be above $1.5 M_{\odot}$ for $U_\Delta (n_0)>-109$~MeV, cf. left panel of Fig.~\ref{fig:cut03-Udep-1}. The maximum NS mass in the KVORcut03$\Delta$ model for $U_\Delta (n_0) =-50$~MeV is 2.17 $M_{\odot}$ that is only by  $0.01\, M_\odot$ less than  that in the original  KVORcut03 model. It decreases only slightly for  more attractive potentials $U_\Delta $, cf. Fig.~\ref{fig:cut03-Udep-1}, right.
In the KVORcut03H$\Delta\phi\sigma$ model the maximum NS mass is $\simeq 2.08\, M_\odot$ and in the KVORcut03H$\Delta\phi$ model $\simeq 1.97\,M_\odot$, being in both cases almost independent on the value of $U_\Delta$. Thus, even for such an unrealistically attractive potential $U_\Delta (n_0)= -150$\,MeV, the maximum mass constraint remains satisfied (although marginally for the KVORcut03H$\phi$, KVORcut03H$\Delta\phi$  models), cf. Figs.~\ref{fig:cut03-Mn} and \ref{fig:cut03-MR}. The NS radius changes only slightly (by less than 0.5\,km) even for $U_\Delta =-150$\,MeV in the KVORcut03H$\phi$, KVORcut03H$\Delta\phi$  models.

It proved to be that within the MKVOR$\Delta$ model in ISM  the nucleon effective mass $m_N^*$   vanishes at $n=n_{{\rm c},f=1}$, cf. Fig. \ref{fig:mkv-meff} (e.g., $n_{{\rm c},f=1}\simeq 5.8\,n_0$ for $U_\Delta (n_0)=-50$~MeV).
Thus in the given model the hadron EoS should be unavoidably replaced to the quark one for  higher densities.  To extend application of a hadronic model to densities $n>n_{{\rm c},f=1}$ we formulated a modification of the MKVOR model, which introduces the cut mechanism  both in the $\om$ and $\rho$ sectors. We label it as the MKVOR* model, see scaling functions and $f(n)$ in Figs.~\ref{Fig-1-new} and \ref{fig:eta_r}, respectively.
The MKVOR* model differs from the MKVOR model in the scaling function in the $\om$ sector only for large values of the scalar field, $0.95<f<1$, that corresponds to densities $n\gsim 5n_0$. This limits a decrease of the nucleon effective mass in the ISM. For BEM $f\lsim 0.6$ and results for MKVOR*-based models coincide with those for the corresponding MKVOR-based models.

The MKVOR* model is more sensitive to inclusion of $\Delta$s than KVORcut03 model since
in the former model the effective nucleon mass is smaller.
In the MKVOR*$\Delta$ model, as in MKVOR one, the effective nucleon mass in ISM demonstrates a back-bending behaviour in some density region provided $U_\Delta$ is chosen to be more attractive than $-67$\,MeV. For $U_\Delta >-67$\,MeV the effective nucleon mass decreases monotonously with an increase of the density, cf. Fig.~\ref{fig:mkvstar-meff} (left). The $\Delta$ concentration  demonstrates a similar behavior,  cf. Fig.~\ref{fig:mkvstar-meff} (middle).
The pressure as a function of the density in ISM,  cf. Fig.~\ref{fig:mkvstar-meff} (right), for $U_\Delta >-56$\,MeV has a behaviour typical for a third-order phase transition. For $U_\Delta <-56$\,MeV the transition to the state with non-zero $\Delta$ concentration is of the first order. For $-67\,{\rm MeV}<U_\Delta <-56$\,MeV there is one spinodal region, whereas for $U_\Delta <-67$\,MeV the $P(n)$  curve has a back bending in some density interval, and there exist two spinodal regions.
This example is in detail studied, cf. Fig.~\ref{Fig-P-new}. The presence of a first-order phase transition owing to the appearance of $\Delta$s could manifest itself through an increase of a pion yield at typical energies and momenta corresponding to the $\Delta$ decays in heavy-ion collision experiments.

In BEM $\Delta$s appear in the MKVOR*$\Delta$ model already at $n=2.5\,n_0$ for
$U_\Delta = - 50$\,MeV and at $n = 1.7\,n_0$ for $U_\Delta = - 100$\,MeV. In MKVOR*H$\Delta\phi$ and MKVOR*H$\Delta\phi\sigma$   models $\Delta$s appear at smaller densities than hyperons but
their presence does not substantially change the NS compositions compared
with the case without $\Delta$s. The critical densities of $\Lambda$ and
$\Xi^-$ hyperons increase with a decrease of $U_\Delta$, opposite to that occurs for the
concentration of $\Xi^0$. For $U_\Delta = -50$\,MeV, $\Xi^0$ do not arise, cf. Figs.~\ref{fig:mkv-conc} and \ref{fig:mkv-Udep-nc}. Despite the presence of $\Delta$s  affects substantially  the NS composition, the star mass changes rather weakly, cf. Fig.~\ref{fig:mkv-Mn}. For a realistic value of the $\Delta$ potential, $U_\Delta=-50$\,MeV, the NS mass decrease proves to be tiny. For a deep $\Delta$ potential, $U_\Delta=-100$\,MeV, change of the NS mass does not exceed $0.2\,M_\odot$. The maximum NS mass changes even smaller (by $\lsim 0.05$\,$M_\odot$) so that the maximum mass constraint is safely fulfilled even after the inclusion of both $\Delta$ baryons and hyperons. The DU constraint $M_{\rm DU} > 1.5\,M_{\odot}$ proves to be fulfilled for
$U_\Delta >- 88$\,MeV, cf. Fig.~\ref{fig:mkv-Udep-nd-mm} (left panel). The maximum mass of the NS decreases only slightly with a deepening of $U_\Delta$ and  remains substantially larger than the maximum measured pulsar mass ($2.01 \pm 0.04M_\odot$ for PSR J0348+0432), cf. Fig. \ref{fig:mkv-Udep-nd-mm}, right. Inclusion of $\Delta$s in MKVOR-based models with or without hyperons does not change noticeably the mass-radius relation for NSs for $U_\Delta=-50$\,MeV. For $U_\Delta=-100$\,MeV the radius of the NS with the mass $1.5\,M_\odot$ decreases by $\sim $ 0.5\,km, cf. Fig.~\ref{fig:mkv-Udep-mr}.

Concluding, we included $\Delta$ isobars in the RMF models with scaled effective hadron masses and couplings. We demonstrated that for reasonable values of the $\Delta$ potential (in the range of $-(50\mbox{--100})$\,MeV) and for the ratios of the coupling constants given by SU(6) model ($x_{\om\Delta}=x_{\rho\Delta}$=1, see Eq.~(\ref{x-QCU})) the KVORcut03$\Delta$-based and MKVOR*$\Delta$-based models appropriately satisfy the constraints considered previously in~\cite{Maslov:2015msa,Maslov:2015wba} within the KVORcut-based and MKVOR-based models with and without hyperons, excluding $\Delta$ isobars.
Thus, we demonstrated that within our models the $\Delta$ puzzle is resolved as well as the hyperon puzzle.

\section*{Acknowledgement}
We thank M.~Borisov and F.~Smirnov for the interest in this work. The reported study was funded by the Russian Foundation for Basic Research (RFBR) according to the research project No 16-02-00023-A. The work was also supported
by the Slovak Grant No. VEGA-1/0469/15,   by ``NewCompStar'', COST
Action MP1304 and by the Ministry of Education and Science of the Russian Federation (Basic part). Computing was partially performed in the High Performance Computing Center of the Matej Bel University using the HPC infrastructure acquired in Project ITMS 26230120002 and 26210120002 (Slovak infrastructure for high-performance computing) supported by the Research \& Development Operational Programme funded by the ERDF.  E.E.K. thanks the Laboratory of Theoretical  Physics at JINR (Dubna) for warm hospitality and acknowledges the support by grant of the Plenipotentiary of the Slovak Government to JINR.

\end{document}